\documentclass[a4paper,11pt]{article}
\usepackage[utf8]{inputenc}
\usepackage[margin=0.9in]{geometry}
\usepackage{multicol}
\usepackage{mathtools}
\usepackage{enumitem}
\usepackage{bm}
\usepackage{float}
\usepackage[framemethod=tikz]{mdframed}
\usepackage{lipsum}
\usepackage[toc,page]{appendix}
\usepackage{graphics}
\usepackage[utf8]{inputenc}
\usepackage[english]{babel}
\usepackage{hyperref}
\usepackage{float}
\usepackage{graphicx}
\usepackage{wrapfig}
\usepackage{pgf} 
\usepackage[utf8]{inputenc}
\usepackage{graphicx}
\usepackage{comment}
\usepackage{lscape} 
\usepackage{framed}
 \usepackage{multirow}
    \usepackage{amsmath}
    \usepackage{subcaption}
	\usepackage{listings}
\usepackage{soul}
    
    \usepackage{graphicx}
    \usepackage{dcolumn}
    \usepackage{bm}
    \usepackage{amsmath} 
\usepackage{colortbl}
\newcolumntype{L}[1]{>{\raggedright\let\newline\\\arraybackslash\hspace{0pt}}m{#1}}
\usepackage[toc,page]{appendix}

\hypersetup{
  colorlinks,
  allcolors=.,
  urlcolor=blue,
}

\usepackage{amsmath}
\begin{document}

\title{ Bayesian Cluster Geographically Weighted Regression for Spatial Heterogeneous Data}

\author{
Wala Draidi Areed, Aiden Price, Helen Thompson,  Conor Hassan, Reid Malseed,  Kerrie Mengersen}


\maketitle
\begin{abstract}
Spatial statistical models are commonly used in geographical scenarios to ensure spatial variation is captured effectively. However, spatial models and cluster algorithms can be complicated and expensive. This paper pursues three main objectives. First, it introduces covariate effect clustering by integrating a Bayesian Geographically Weighted Regression (BGWR) with a  Gaussian mixture model and the Dirichlet process mixture model. Second, this paper examines situations in which a particular covariate holds significant importance in one region but not in another in the Bayesian framework. Lastly, it addresses computational challenges present in existing  BGWR, leading to notable enhancements in Markov chain Monte Carlo estimation suitable for large spatial datasets. The efficacy of the proposed method is demonstrated using simulated data and is further validated in a case study examining children's development domains in Queensland, Australia, using data provided by Children's Health Queensland and Australia's Early Development Census.
\end{abstract}



\maketitle
\section{Introduction}
 Spatial regression models and algorithms are widely used to model the relationships between the response variable and the covariates over the region of interest. Cressie \cite{cressie2015statistics}
proposed a spatial linear regression model in which only the intercept accounts for the spatial random effect. Diggle \cite{diggle1998model} extended the spatial linear regression to the spatial generalised linear model. Brundson \cite{brunsdon1996geographically} proposed geographically weighted regression (GWR) to capture smoothly varying patterns of the regression coefficients. The GWR fits a local weighted regression model at the location of each observation and captures spatial information by accounting for nearby observations, using a weight matrix defined by a kernel function. Xue \cite{xue2020geographically} extended the GWR to the Cox survival model, by providing a novel approach to analysing spatially dependent survival data. This extension enhances the ability to explore how geographic factors impact time-to-event outcomes. 

Despite the appeal of the GWR model, there are some limitations in these frequentist approaches. For example, a critical issue is the violation of the usual assumption of non-constant variation between observations, and the resultant normality assumption for the errors \cite{chan2008incorporating}. Additionally, it struggles to address issues of model complexity, overfitting, variable selection and  multicollinearity; also, the stability and reliability of frequentist GWR  might yield unstable results or high variance when dealing with small sample sizes \cite{f2007methods}. Bayesian GWR is considered one of the best
solutions to address these problems \cite{sodikin2017geographically}. In the Bayesian framework, Gelfand \cite{gelfand2016spatial} built a model with spatially varying coefficients by applying a Gaussian process to the distribution of regression coefficients. LeSage \cite{lesage2004family} suggested an early version of  BGWR, where the prior distribution of the parameters depends on expert knowledge. More recent approaches have been proposed by  Ma \cite{ma2021geographically}, who proposed BGWR based on the weighted log-likelihood, and Liu \cite{liu2021generalized} proposed BGWR based on a weighted least-squares approach. 

In this paper, we propose a new extension that integrates BGWR with unsupervised probabilistic clustering. Cluster analysis is of great interest in many spatial contexts. The most common method for spatial clustering is the scan statistic \cite{kulldorff1995spatial}, which is constructed via likelihood ratio statistics. Similar efforts have been made in Bayesian and non-parametric frameworks. For example, Li \cite{li2015bayesian} proposed a non-parametric Bayesian method to find cluster boundaries for areal data. Neill and Moore \cite{neill2005detection} described an extension of the spatial scan statistics based on improving space-time cluster detection. 
Unsupervised clustering is a set of statistical and machine learning approaches that partition cohorts into subgroups based on the structure of the data. The most common unsupervised clustering algorithms are $K$-means \cite{wu2012cluster}  and a Gaussian mixture model (GMM) \cite{mclachlan1988mixture}. We extend the BGWR  to identify groups of observations that exhibit similar behaviour, by clustering the posterior regression coefficients obtained from BGWR with a Gaussian mixture model and the Dirichlet process mixture model. The proposed BGWR model thus clusters the coefficients into distinct homogeneous groups using these two probabilistic clustering algorithms. One notable advantage of our algorithm is its ability to automatically determine the optimal number of clusters without requiring prior knowledge, which sets it apart from the traditional $K$-means algorithm and provides an edge over frequentist GWR.

In this paper, we also introduce methods that significantly improve the Markov chain Monte Carlo (MCMC) estimation for the BGWR model proposed by Ma \cite{ma2021geographically}. Due to the high computational cost of their algorithm, the geographic regions must be divided into subsets and compute BGWR for each subset separately. This compromised the accuracy of the model for areas on the boundaries of the subsets. In our proposed approach this is not necessary, making it more suitable for large spatial scales and numerous areas while preserving information from the boundaries. Our approach involves removing unnecessary computations and factorizing declarations. It also enables the inclusion and efficient handling of both continuous and categorical explanatory variables. We also integrate BGWR with dynamic variable selection using the reversible jump Markov chain Monte Carlo(RJMCMC), which identifies when a particular covariate is important in a specific local region. 

The power of the proposed algorithm is demonstrated through comprehensive simulation studies. As a practical illustration, this methodology is then applied to a case study focusing on children's development domains in Queensland, Australia.

Child development includes the biological, psychological, and emotional changes that occur between birth and maturity \cite{unicef1993facts}.  Children's development in the early years from birth to five years of age is crucial since it is at this time that the foundations for health development, emotional wellbeing, and life success are built \cite{hertzman1996child}. The early identification of groups of children who are developmentally vulnerable may lead to prompt early intervention. Physical, social, emotional, speech and language, and communication skills are the five critical domains of growth \cite{irwin2007early}. Development domains have been used in previous research to classify children into subgroups that describe their development using an unsupervised clustering algorithm \cite{areed2022vulnerable,geluk2014identifying,ukoumunne2012profiles}. 
While BGWR offers a more robust framework for spatial regression, there remains a challenge in identifying and interpreting localized patterns or homogeneity within the spatial data. This is where the integration with clustering becomes important. Clustering, especially with advanced methods like Gaussian mixture models and the Dirichlet process mixture model, allows us to group similar spatial behaviours  based on the posterior regression coefficients derived from BGWR. We are aiming to identify regions  within the spatial data where specific behaviours or patterns are consistent, enabling targeted interventions or insights.

The novelty of our approach lies in this unique integration. By combining BGWR with probabilistic clustering, we not only get a refined understanding of spatial relationships but also connect these relationships into distinct clusters or groups. This analysis offers depth via BGWR and comprehensive insights through clustering.  Moreover, our methodology's capability to autonomously determine the optimal number of clusters offers a more adaptive and intuitive clustering mechanism than traditional methods. In the context of children's development domains, this can be valuable. Identifying clusters can help in pinpointing areas or groups of children who might have similar developmental challenges or needs, leading to more targeted interventions and policy implementations.
\section{Methods}
In this section, we introduce the BGWR model and detail the three methodological contributions of this paper. First, we describe the vectorisation method to enhance BGWR efficiency. Second, we extend the BGWR analysis to include variable selection for each covariate in each specific region. Finally, we describe the extension of the BGWR model to incorporate the Dirichlet process mixture model (DPMM) and Gaussian mixture model (GMM) in order to identify clusters of coefficients of interest. 
\subsection{Bayesian geographically weighted regression}
A common assumption for the GWR is that the error terms are normally distributed with constant variance $\epsilon(s) \sim N(0,\sigma^2 I)$ \cite{leung2000statistical} for a specific geographical location $s$. There are many situations in different fields where the assumption of constant variance is invalid. According to Paez et al. \cite{paez2002general}, the error term can be written as $\epsilon(s) \sim N(0,\Omega(s))$, where $\Omega(s)=\sigma^2(s) W(s)$  with $W(s)$ as a diagonal matrix of geographic weights function $f(d_i(s)|b)$ that is a decreasing
function of distance $d_i(s)$ from the location $s$ to the location $i$. GWR is seen as a locally
weighted regression method that operates by assigning a weight to each and every
observation $i$ depending on its distance from a specific geographical location $s$ \cite{wheeler2005multicollinearity}.
This local perspective of the variance is often called location heterogeneity. 
 A common approach is to define the observations that are within a certain distance $b$ from a specific location $s$. Different weights can be used in the GWR model, including:
\[ W(s)=\begin{cases} 
      1 & d_i(s)  \leq b \\
      0 & \text{otherwise}
   \end{cases}
\]
\noindent where $d_i(s)$ is the distance between the locations $s$ and $i$. Other weights include the exponential and Gaussian functions, which give the following expressions for weights, respectively:
\begin{equation}
    W(s)=e^ {-(d_i(s)/b)}     \label{expo}
\end{equation}
\begin{equation}
     W(s)=e^{-(d_i(s)/ b)^2} \label{gus}
\end{equation}

\noindent where $b$ represents the bandwidth that controls the decay over distance \cite{cho2010geographically}. Equations (\ref{expo}) and  (\ref{gus}) are decreasing functions of $d_i(s)$, which shows that an observation far away from the location of interest contributes little to the estimate of parameters at that location. Different choices have been used to find $d_i(s)$; the Euclidean distance is the most popular choice when latitude and longitude for each observation are available \cite{yu2020inference}. Other choices of distance matrices include the graph distance \cite{gao2010survey} and greater circle distance (GCD)\cite{carter2002great}; these approaches are used in the BGWR in sections \ref{simulated} and \ref{real}. A further explanation of these distance measures can be found in Appendix \ref{distances}.\\ 
\\
 The likelihood of the BGWR model can be written as \cite{ma2021geographically}:
\begin{equation}
    Y|\beta(s),X,W(s),\sigma^2(s) \sim MVN (X\beta(s),\sigma^2(s) W^{-1}(s))
\end{equation}
$Y$ is the $n \times 1 $ observation or response, $X$ the $ n \times p$  predictor matrix, $\beta$ is $p \times 1 $  vector of spatially varying coefficients, and $\sigma^2(s) W^{-1}(s)$ is an $n \times n$ matrix. Common conjugate prior distributions are $\beta(s)|\sigma^2_\beta \sim N(0,\sigma^2_\beta)$ and $\sigma^2(s) \sim IGamma(\alpha_1,\alpha_2)$. The posterior distribution is given as
\begin{equation}
    p(\beta(s), \sigma^2(s)Y, X,W(s)) \propto p(Y\beta(s),X,W(s),\sigma^2(s))\times (\beta(s) \sigma^2_\beta) \times p(\sigma^2(s))
\end{equation}
 In the GWR model, it is crucial to choose a suitable bandwidth for the weighted function. In a BGWR context, a prior can also be applied to the bandwidth $b$ to allow estimation of the best bandwidth given other parameters. The choice of the prior also depends on the measure of distance that is used. A common prior for the bandwidth is \cite{perez2015selection}:
\begin{equation}
     b \sim Uniform (0,D)   \quad \quad D>0 .
\end{equation}
Without any prior knowledge, $D$ can be selected to be large enough that we begin to approximate with a non-informative prior; i.e., we begin with an approximate global model in which all observations are weighted equally.
\subsection{Vectorisation methodology}
In this section, we introduce our first contribution aimed at enhancing the computational efficiency of the BGWR to accommodate large-scale datasets,  ensure that there is no loss of information at the boundaries, and remove the need to split the data into smaller regions to run the BGWR model.

The use of a multivariate normal distribution in BGWR leads to high computational costs. In this paper, we introduce a new approach based on vectorisation, which is achievable in the "nimble" package in R. This improves efficiency by reducing the number of calculations and nodes in the model and thus enhancing MCMC performance. The likelihoods for each region in the BGWR model are vectorized. The precision matrix $\sigma^2(s) W^{-1}(s)$ is a diagonal matrix, which allows for an alternative representation using univariate Gaussian functions. This permits independent estimation of mean and variance in each dimension and characterization of the multivariate density function as a product of univariate Gaussian functions for each location $s$. 
Given the likelihood function in the multivariate case with a diagonal $W(s)$ for the region $s$, the likelihood becomes:
\begin{align}
f(y(s)|&\beta(s),x(s),W(s),\sigma^2(s)) \nonumber \\
&= \pi^{-1/2} \cdot \sigma^{-1}(s) \cdot |W(s)|^{-1/2} \nonumber \\
&\times \exp\left(-\frac{1}{2} \sigma^{-2}(s) \cdot (y(s) - x^T(s) \beta(s))^T W^{-1}(s) \cdot (y(s) - x^T(s)\beta(s))\right)
\end{align}

\noindent When $W(s)$ is diagonal, the inverse $W^{-1}(s)$ is also diagonal, so the likelihood function can be rewritten as follows:
\begin{align}
f(y(s)|\beta(s),x(s),W(s)) &= \pi^{-1/2} \cdot \sigma^{-1}(s) \cdot |W(s)|^{-1/2} \nonumber \\
&\times \exp\Bigg(-\frac{1}{2} \sigma^{-2}(s) \bigg( (y(s) - x^T(s)\beta(s))^T \nonumber \\
&\quad \times \text{diag}\left(w_1^{-1}(s), w_2^{-1}(s), \ldots,w_n^{-1}(s)\right) \cdot (y(s) - x^T(s)\beta(s)) \bigg) \Bigg)
\end{align}

\noindent Expanding the multiplication term:
\begin{align}
(y(s) - x^T(s)\beta(s))^T \cdot \text{diag}\left(w_1^{-1}(s), w_2^{-1}(s), \ldots,w_n^{-1}(s)\right) \cdot (y(s) - x^T(s)\beta(s)) \nonumber \\
= \sum_{i=1}^n w_i^{-1}(s) \cdot (y(s) - x^T(s)\beta(s))^2.
\end{align}
The likelihood function can then be further simplified to the product of univariate Gaussian likelihoods:
\begin{align}
f(y(s)|\beta(s),x(s),W(s),\sigma^2(s)) &= \pi^{-1/2} \cdot \sigma^{-1}(s) \cdot |W(s)|^{-1/2} \cdot \notag \\
&\quad \exp\left(-\frac{1}{2} \sigma^{-2}(s) \sum_{i=1}^{n} w_i^{-1}(s) \cdot (y(s) - x^T(s)\beta(s))^2\right)
\end{align}
where $\left(w_1^{-1}(s), w_2^{-1}(s), \ldots,w_n^{-1}(s)\right)$ is the diagonal that represents the weights from a specific region $s$ for the $n$ different locations of the weight matrix $W(s)$ for location $s$. Thus, observations with higher weights (indicating higher importance or reliability) contribute more to the overall likelihood value compared to those with lower weights.

The exponent term inside the exponential represents the sum of squared differences between the observed response variable $y(s)$ and the predicted values $x^T(s) \beta$, weighted by $w_i^{-1}$, which is the precision or weight associated with each observation. The determinant of a diagonal matrix is simply the product of its diagonal entries. Thus,
\[|W(s)| = w_1(s) \cdot w_2(s) \cdot \ldots \cdot w_n(s) = \prod_{i=1}^{n} w_i(s)\]
The likelihood in the above form demonstrates that, when the weighted matrix is diagonal, the multivariate Gaussian likelihood reduces to a product of $n$ univariate Gaussian likelihoods, one for each region $s$. 
We can represent the likelihood for each \( y(s) \) as:

\begin{equation}
y(s)|\beta(s), x(s), w_i(s), \sigma^2(s) \sim {N}(x^T(s)\beta(s), \sigma^2(s) w_i^{-1}(s))
\end{equation}

\noindent The full likelihood for the entire dataset is the product of the likelihood for each observation.

In summary, the BGWR approach that is developed in this paper can be represented in the following form:
\begin{equation}
\label{eq:2}
  y(s)|\beta(s), x(s), w_i(i), \sigma^2(s) \sim {N}(x^T(s)\beta(s), \sigma^2(s) w_i^{-1}(s))
\end{equation}
\begin{equation}
\label{q:1}
    \beta_j(s)|\sigma^2_\beta \sim N(0,\sigma^2_\beta)
\end{equation}
\begin{equation}
\label{q:2}
    \sigma^2_\beta \sim IGamma(\alpha,\beta)
\end{equation}
\begin{equation}
\label{q:3}
     \sigma^2(s) \sim IGamma(\alpha_1,\alpha_2)
\end{equation}
\begin{equation}
\label{q:4}
     w_i(s) = f(d_i(s)|b)
\end{equation}
\begin{equation}
\label{q:555}
     b \sim Uniform (0, D)
\end{equation}
\noindent where $f$ is the weighted function in equations \ref{expo} or \ref{gus}, and $b$ is the bandwidth.\\
 The posterior distribution is obtained using MCMC methods and the parameters of interest are estimated using the posterior mean of the respective marginal posterior distributions.  The steps of fitting the proposed model in Nimble \cite{de2017programming} are provided in Appendix \ref{real_nimble}.\\

\subsection{BGWR with Dynamic Variable Selection: RJMCMC Approach} \label{algorithm2}
This section introduces our second contribution, the identification of locally important covariates for each location.

In the Bayesian framework, various algorithms have been developed to determine the most relevant predictors that should be included in a model to provide the best explanation for a response variable. A popular approach was developed by Mitchell and Beauchamp \cite{mitchell1988bayesian}, which employs a prior distribution that encourages sparse models in Bayesian linear regression. Spike and slab approach to variable selection have also been proposed by George and McCulloch, Chipman \cite{chipman1996bayesian}, and Kuo and Mallick \cite{kuo1998variable} proposed a simpler approach that embeds indicator variables in the regression equation.  Green \cite{green1995reversible} introduced RJMCMC, which enables models of different dimensionality to be explored. Bhattacharya and Dunson \cite{bhattacharya2011sparse} proposed a Bayesian non-parametric approach to sparse regression that can handle infinite potential predictors. The approach uses a spike-and-slab prior to calculating posterior model probabilities for each possible subset of variables but does not require specifying the number of potential predictors a priori. Ma \cite{ma2021geographically} employed a traditional spike and slab approach in the context of BGWR. However, their method includes or excludes the coefficient entirely without considering its potential importance in specific regions.

In this paper, a novel approach is described for spatial local BGWR  that leverages the inclusion or exclusion of the coefficients based on their importance in specific regions.
This advancement enhances the model's ability to capture location-specific effects accurately and opens new possibilities for understanding the spatially varying impact of coefficients.
The proposed algorithm replaces the likelihood in equation (\ref{eq:2}) as: 
\begin{equation}
\label{RJJ}
 y(s)|\beta(s), x(s), w_i(s), \sigma^2(s) \sim {N}(x^T(s)( \Gamma_{s} * \beta(s), \sigma^2(s) w_i^{-1}(s))
 \end{equation}
    \begin{equation}
    \label{ber}
       \gamma_{j}(s) \sim Bernoulli(\psi_j) 
    \end{equation}
        \begin{equation}
    \label{bet}
      \psi_j \sim  Beta(1,1)
    \end{equation}
    where  $\Gamma_{s}$  is a $1 \times p$ vector with elements $\gamma_{j}(s)$, and $*$ presents the broadcast operation (element wise multiplication),
with other priors given by equations (\ref{q:1}) to (\ref{q:555}).
In equation (\ref{RJJ}), for each region $s$ and each corresponding covariate $j$, an indicator variable $\gamma_{j}(s)$ is assigned, following a Bernoulli distribution with probability parameter $\psi_j$. Thus $\gamma_{j}(s)$ allow probabilistic determination of whether the coefficient $\beta_j$ should be included in the model for the region $s$ or not, depending on the contribution of the $j$th covariate in the model for the $s$th region. This dynamic approach to variable selection allows the model to make data-driven decisions about the importance of covariates in explaining the response variable for each region. A RJMCMC algorithm is adopted to implement this approach \cite{green1995reversible}.
The steps of the RJMCMC algorithm are detailed in Appendix \ref{RJMCMCapp}. We have tested our algorithm using simulated and real datasets; the code for the analysis of the simulated data is available on the first author's GitHub. \href{https://github.com/waladraidi/BGWR-Clustering}{Link to BGWR Clustering on GitHub}

\subsection{Cluster BGWR}
In this section, we present two techniques for probabilistic clustering of the posterior distributions of the region-specific (local) coefficients obtained from the BGWR analysis. The first parametric unsupervised learning technique assumes that the coefficients can be represented by a known probability distribution, the Gaussian mixture model (GMM). The second is a non-parametric unsupervised learning method that does not make any assumptions about the probability distribution of the coefficients and aims to discover the clusters, using a Dirichlet process mixture model (DPMM).  We describe these two approaches in a univariate context and consider a random sample of size $n$ drawn from the posterior distribution of the coefficient of interest, obtained from the MCMC iterations. We denote this sample generically as $y=\{y_1,y_2,...,y_n\}$.

\subsection*{Gaussian mixture model}
 The GMM is represented as a weighted sum of Gaussian density functions, each with its own set of parameters \cite{dempster1977maximum}.
 The GMM is represented as
\begin{equation}
    p(y|\theta)=\sum_k^K \alpha_k p_k(y_i|\theta_k)
\end{equation}
where $\sum_{k}^{K} \alpha_k=1$ and $p_k(y_i|\theta_k)$ is a Gaussian density function parameterized by $\theta_k$ \cite{he2010laplacian}. For computational simplicity and to enhance inferential capability, a latent component indicator, $z=\{z_1,z_2,..,z_n\}$ is introduced, where each $z_i$ is a $K$-dimensional vector and the $k$th element of $z_i$ takes the value 1 if $y_i$ comes from component $k$, $(k=1,..., K)$ and takes the value 0 otherwise.\\
The parameters of the GMM can be estimated in a Bayesian framework using MCMC or approximations such as variational inference. Another popular approach that we adopted here is expectation-maximization (EM) \cite{he2010laplacian}. The EM algorithm iterates between two steps until convergence: the expectation step, which takes the conditional expectation of the complete data log-likelihood given the observed data and current parameter estimates, and the maximization step, which maximizes the log-likelihood with respect to the parameters to give updated estimates \cite{reynolds2009gaussian}.
\\There are many ways to select the optimal number of clusters ($K$) for GMM.  Common approaches include the silhouette score \cite{rousseeuw1987silhouettes}, the Bayesian information criterion (BIC) \cite{wan2019novel}, the elbow method which is based on a plot of the explained variance of the model versus the number of clusters \cite{liu2017novel}, and cross-validation, which involves splitting the data into multiple subsets and then training and evaluating the GMM model on different subsets of the data \cite{fu2020estimating}. In this study, we used BIC to find the optimal number of clusters. The choice of BIC is motivated by the nature and assumptions of the GMM. The GMM assumes that data are generated from a mixture of several Gaussian distributions. BIC is particularly suited for model selection in probabilistic models like GMM. It balances the likelihood of the data under the model against the complexity of the model, effectively penalizing overfitting. Given that GMM involves the estimation of several parameters, especially as the number of mixtures (or clusters) increases, BIC becomes a particularly relevant criterion for determining the optimal model.
\subsection*{Dirichlet process mixture model}
DPMM is a non-parametric method that replaces the fixed number of clusters with a random probability measure \cite{ferguson1973bayesian}. The DPMM is defined by a base distribution $G_0$ and a concentration parameter $\alpha$. The base distribution represents the prior distribution over the means of the clusters, while the concentration parameter controls the level of clustering.

\noindent The DPMM can be summarised as follows:
\begin{equation}
    y_i|\theta_i \sim p(y_i|\theta_i)
\end{equation}
\begin{equation}
   \theta_i|G \sim G
\end{equation}
\begin{equation}
   G \sim DP(\alpha,G_0)
\end{equation}
where each $\theta_i$ is drawn from a mixing distribution $G$. This mixing distribution has a Dirichlet process prior, with concentration parameter $\alpha$ and base distribution $G_0$ where $E[G]=G_0$ \cite{sammut2011encyclopedia}. The concentration parameter acts as an inverse variance where larger values of
$\alpha$ result in smaller variances, which creates more concentrated draws around the mean of the
base distribution \cite{teh2010dirichlet}. 
The DPMM is typically estimated using a Bayesian approach, such as the Chinese restaurant process or the stick-breaking process \cite{neal2000markov}. The stick-breaking construction is used in this paper. In this construction, the $G$ can be determined by an infinite sum of weighted point masses:
\begin{equation}
    G=\sum_{k=1}^{\infty} C_k \delta_{\theta_k}
\end{equation}
where $\delta_{\theta_k}$ is a point mass of 1 located at $\theta_k$, which is sampled directly from $G_0$, i.e $\theta_k \sim G_0$.The weights $C_k$ are obtained through the stick-breaking process:
\begin{align}
\begin{split}\label{eq:1}
    & V_1,V_2,... \sim Beta(1,\alpha)\\
         & C_1=V_1\\
          & C_k=V_k \prod_{j=1}^{k-1} (1-V_j);  \quad \quad  k\ge 2.
\end{split}
\end{align}

\subsection*{Cluster configurations}
 Two approaches were considered to determine cluster membership of the observed data. The first method, known as Dahl's approach \cite{dahl2006model}, involves computing the membership matrices for each iteration, denoted as \(B^{(1)}, \ldots, B^{(M)}\), with  \(M\) being the number of post-burn-in MCMC iterations. The membership matrix for the \(c\)th iteration \(B^{(c)}\) is defined as:
 \begin{equation}
     B^{(c)} = (B^{(c)}(i, j))_{i,j \in \{1:n\}} = \mathbf{1}(z^{(c)}_i = z^{(c)}_j)_{n \times n}
 \end{equation}
where $\mathbf{1}(\cdot)$ represents the indicator function. The entries $B^{(c)}(i, j)$ take values in $\{0, 1\}$ for all $i, j = 1, \ldots, n$ and $c = 1, \ldots, M$ with $B^{(c)}(i, j) = 1$, indicating that observations $i$ and $j$ belong to the same cluster in the $c$th iteration.
An empirical estimate of the probability that locations $i$ and $j$ are in the same cluster is given by the average of $B^{(1)}, \ldots, B^{(M)}$:
\begin{equation}
   \overline{B} = \frac{1}{M} \sum_{c=1}^M B^{(c)}
\end{equation}
where $\sum$ denotes the element-wise summation of matrices. The $(i, j)$th entry of $\overline{B}$ provides the required empirical estimate.\\
Subsequently, we determine the iteration that exhibits the least squared distance to $\overline{B}$ as:

\begin{equation}
   C_{LS} = \arg \min_{c \in \{1:M\}} \left[\sum_{i=1}^n \sum_{j=1}^n (B^{(c)}(i, j) - \overline{B}(i, j))^2\right]
\end{equation}

The second method utilised here is the mode method. This method leverages the posterior samples  of $z_i$, where $z$ denotes the cluster assignments specific to each region.
Each iteration generates a new set of cluster assignments $z$, which are dependent on the parameters. Consequently, following multiple iterations, each region will have a collection of cluster assignments $z$. The mode indicates the cluster with the highest probability of assignment for a given region. Moreover, obtaining probabilities for assignment to alternative clusters provides valuable insights, aiding in the inferential and decision-making process.

\subsection*{Cluster accuracy}
To assess the accuracy of our proposed algorithm, we utilised the Rand  index (RI) \cite{rand1971objective} in our simulated study (see section \ref{simulated}), to perform a comparison between the cluster configurations obtained  using either Dahl's method or the mode method, and the true clusters. Specifically, we employed this metric for the simulated data. However, for real data analysis, true labelling is unavailable. The RI quantifies the level of agreement between two sets of cluster assignments, denoted as \(C\) and \(C'\), with respect to a given dataset \(X = \{x_1, x_2, \ldots, x_n\}\). Each data point  \(x(s)\) is assigned a cluster label \(c_i\) in  \(C\) and \(c'_i\) in \(C'\).
The computation of RI is based on the following formula:
\[
\text{RI} = \frac{a + b}{a + b + c + d}
\]
where \(a\) is the number of pairs of data points that are in the same cluster in both \(C\) and \(C'\) (true positives), \(b\) is the number of pairs of data points that are in different clusters in both \(C\) and \(C'\) (true negatives), \(c\) is the number of pairs of data points that are in the same cluster in \(C\) but in different clusters in \(C'\) (false positives), \(d\) is the number of pairs of data points that are in different clusters in \(C\) but in the same cluster in \(C'\) (false negatives).

The RI value ranges from 0 to 1, where a value of 1 signifies a perfect agreement between the two clusterings (both \(C\) and\(C'\) fully agree on all pairs of data points). On the other hand, a value close to 0 indicates a low level of agreement between the two clusterings.
\section{Simulated Data Analysis} \label{simulated}
In this section, we present two simulated datasets: one with spatially varying coefficients and the other without. The first simulation is related to a dataset based on Louisiana and is generated in a manner similar to the work by Ma \cite{ma2021geographically}. This dataset comprises a total of 64 regions, with three observations in each region $i$. 

Under the scenario in which there are no spatially varying coefficients, we generated independent continuous covariates denoted as $(X_1, X_2, ..., X_5)$ from a standard normal distribution $N(0, 1)$ for each region. These covariates are used to create the response vector $Y$, generated as $Y = X\beta + \epsilon$, where $\epsilon \sim N(0, 1)$. The parameter vector $\beta$ is set to be $\beta = (2, 0, 0, 4, 8)$.

We used a bandwidth parameter $D$ set to 100. The maximum great circle distance in the spatial structure of the 64 regions is 10, so using a bandwidth of 100 induces a weighting scheme that ensures relative weights are assigned appropriately. If the distance between two regions is considerable, the relative weight is approximately $exp(-10/100)= 0.904$. This approximation thus allows the model to behave similarly to a global model where every observation is equally weighted, ensuring a sufficiently non-informative prior bandwidth $b$.

In each region, three observations are generated, resulting in a total of 192 observations per replicate. The analysis is replicated 100 times. Each replicate involves running a MCMC chain of length 10,000 without thinning, and the initial 2,000 samples are discarded as burn-in.
The mean bandwidth selected in the 100 replicates was calculated as well as the posterior means of the parameters $\beta$. The results are reported in Table \ref{tab:my-table1}. Additionally, we employed the dynamic variable selection (RJMCMC) technique described in section \ref{algorithm2}. This analysis wasreplicated 100 times. Each replicate involves running a RMCMC chain of length 400,000 without thinning, and the initial 20,000 samples are discarded as burn-in to converge. The results are summarized in Table \ref{tab:my-table2}. This table illustrates how our algorithm identifies important covariates for each location. 

\begin{table}[ht!] 
\caption{ \label{tab:my-table1} Average parameter estimates and their performance when there is no spatial variation in the underlying true parameters. The performance metrics used include mean absolute bias (MAB), mean standard deviation (MSD), and mean of mean squared error (MMSE).}
\centering
\begin{tabular}{llllll}
\hline
        & $\overline{\hat{\beta}}$ & MAB & MSD & MMSE & bandwidth \\ \hline
$\beta_1$ &  1.975    &   0.079   &   0.012 & 0.006     &   49.06       \\
$\beta_2$ &  -0.089    &   0.014  &   0.011  &  0.005    &           \\
$\beta_3$ &  0.074    &   0.012   &  0.076   &   0.007   &           \\
$\beta_4$ & 3.927     & 0.013    & 0.072    &    0.004  &           \\
$\beta_5$ &  8.383    &  0.009   & 0.06    &    0.003  &           \\ \hline
\end{tabular}
\end{table}

\begin{table}[ht!] 
\caption{\label{tab:my-table2} Average parameter estimates and the performance of parameter estimates when there is no spatial variation in the underlying true parameters using dynamic variable selection. The performance metrics used include MAB, MSD, and MMSE.}
\centering
\begin{tabular}{lcccccc}
\hline
& True $\beta$ & $\overline{\hat{\beta}}$ & MAB & MSD & MMSE & bandwidth \\
\hline
$\beta_1$ & 2 & 1.96   &   0.048  &   0.081   &    0.002 &    47.89      \\
$\beta_2$ & 0 & -0.0003   &  0.027   &    0.058 &    0.0007  &           \\
$\beta_3$ & 0 & 0.0001   &  0.018   &   0.046  &  0.0003    &           \\
$\beta_4$ & 4 & 3.92    &  0.014   &  0.066   &   0.002   &           \\
$\beta_5$ & 8 & 7.95    &  0.060  &  0.072   &   0.003    &        \\
\hline
\end{tabular}
\end{table}

\noindent Our proposed BGWR model with vectorisation significantly improves computational efficiency, completing each replicate in under 250 seconds. In contrast, a model that emulated the previous approach suggested by Ma \cite{ma2021geographically}, which relies on a multivariate normal distribution to calculate the likelihood \cite{ma2021geographically}, took over 15 minutes to run for each replicate.  Finally, we evaluated various weighted functions in this analysis. Table \ref{tab:my-table3} presents the WAIC, DIC, and effective sample size ($P_D)$ obtained with different kernels. The explanation of WAIC and DIC for the BGWR model can be found in the Appendix \ref{assessment}. The Bi-square kernel demonstrated a better fit to the data compared to the exponential and Gaussian kernels.

\begin{table}[ht!] 
\caption{\label{tab:my-table3} Model assessment for the proposed algorithm BGWR using different kernels for the simulated data.}
\centering
\begin{tabular}{llll}
\hline
           & Exponential kernel & Bi-square kernel & Gaussian kernel  \\ \hline
WAIC       & 181878.4            & 181854           & 181855          \\ 
DIC        & 35846.3             & 35831.9          & 35853.4         \\
$P_D$      & 391.7               & 381.8            & 404.7           \\ \hline
\end{tabular}
\end{table}

\begin{figure}[ht]
     \centering
     \includegraphics[scale=0.6]{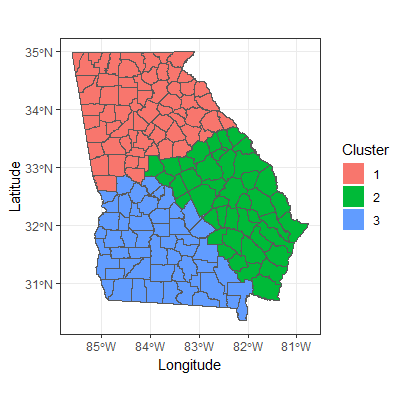}
     \caption{Cluster assignment for Georgia
regions used for simulation studies.}
     \label{fig:enter-labelw}
 \end{figure}

The second simulated study was created based on the structure of the state of Georgia, also considered by Ma \cite{ma2020heterogeneous}. This dataset includes 159 regions. Six spatially correlated covariates ($X_1$ to $X_6$) were generated using multivariate normal distributions with spatial weight matrices derived from the distance matrix and parameter bandwidth. The response variable ($Y$) in the simulation was generated by the GWR model:
\[ y(s) = \beta_{0}(u(s), v(s)) + \sum_{k=1}^{K} \beta_{k}(u(s), v(s)) \cdot X_{k}(s) + \varepsilon(s) .\]
Notably, the parameters ($\beta_1$ to $\beta_6$) of this GWR  model were spatially varying, based on the spatial weight matrices. We then visually partitioned the areas into three large regions to define true clustering settings based on the spatial coordinates of centroids. This approach allowed us to create distinct spatial patterns in the data, which incorporated spatial autocorrelation, spatial variability, and true clustering settings. 
 Figure \ref{fig:enter-labelw} summarises the partition of the Georgia areas into three large regions with sizes 51, 49, and 59 regions in the three clusters, respectively. The same parameter vectors from Table \ref{table_true} were utilised for all three clusters across three settings under different strengths of signals. Setting 1 shows relatively low signal strengths for all three clusters. The signals in this setting are primarily found in specific variables within each cluster, with no variable consistently having a strong signal across all clusters. Setting 2 exhibits stronger signals than Setting 1, with more pronounced variability among the clusters. Cluster 3 in this setting shows the strongest signals among the clusters, and certain variables have high signal strengths across multiple clusters. The signals are well spread across different variables in each cluster, and the magnitude of the signals is higher than in the previous settings.
The performance of the proposed algorithm is presented in Table \ref{tab:my-table 60}. The analysis includes the three settings (Setting 1, Setting 2, and Setting 3) and the two clustering methods (GMM and DPMM). It is apparent that Setting 3 consistently exhibits higher RI values, signifying superior clustering accuracy compared to the other settings. Furthermore, the number of clusters varies across settings and methods. As anticipated, the DPMM method generally generates a larger number of clusters than the GMM method across most settings \cite{liu4346394topic}. Additionally, the cluster configurations differ across settings and methods, revealing distinctive patterns and structures in the data for each configuration. The results for the final clusters in each setting are visually presented in Appendix \ref{sim_further_test}.
\begin{table}[ht!] 
\caption{\label{table_true} True parameter vectors used in data generation for three clusters.}
\centering
\begin{tabular}{ccccccc}
\hline
Setting & Cluster 1 & Cluster 2 & Cluster 3 \\
\hline
1 & (1, 0, 1, 0, 0.5, 2) & (1, 0.7, 0.3, 2, 0, 3) & (2, 1, 0.8, 1, 0, 1) \\
2 & (2, 0, 1, 0, 4, 2) & (1, 0, 3, 2, 0, 3) & (4, 1, 0, 3, 0, 1) \\
3 & (9, 0, -4, 0, 2, 5) & (1, 7, 3, 6, 0, -1) & (2, 0, 6, 1, 7, 0) \\
\hline
\end{tabular}
\end{table}
\begin{table}[H]
\caption{\label{tab:my-table 60} The cluster accuracy and cluster configuration using three different settings when there is spatial variation in the underlying true parameters based on Georgia dataset.
}
\scalebox{0.65}{
\centering
\begin{tabular}{lllllll}
\hline
                                                        & \multicolumn{2}{l}{Setting 1}                                                                                                                                                                      & \multicolumn{2}{l}{Setting 2}                                                                                                                                                   & \multicolumn{2}{l}{Setting 3}                                                                                                                                               \\ \hline
                                                        & GMM                                                          & DP                                                                                                                                  & GMM                                                            & DP                                                                                                             & GMM                                                                                        & DP                                                                             \\ \hline
\multicolumn{1}{l|}{RI for Dhal method}                 & 0.78                                                         & \multicolumn{1}{l|}{0.68}                                                                                                           & 0.80                                                           & \multicolumn{1}{l|}{0.79}                                                                                      & 0.76                                                                                       & 0.84                                                                           \\
\multicolumn{1}{l|}{RI for mode method}                       & 0.76                                                         & \multicolumn{1}{l|}{0.64}                                                                                                           & 0.76                                                           & \multicolumn{1}{l|}{0.82}                                                                                      & 0.85                                                                                       & 0.88                                                                           \\
\multicolumn{1}{l|}{Number of clusters (Dahl's method)} & 3                                                            & \multicolumn{1}{l|}{10}                                                                                                             & 3                                                              & \multicolumn{1}{l|}{7}                                                                                         & 8                                                                                          & 5                                                                              \\
\multicolumn{1}{l|}{Number of clusters mode}            & 3                                                            & \multicolumn{1}{l|}{7}                                                                                                              & 3                                                              & \multicolumn{1}{l|}{4}                                                                                         & 5                                                                                          & 6                                                                              \\
\multicolumn{1}{l|}{Cluster summary (Dahl's method)}     & \begin{tabular}[c]{@{}l@{}}C1=54,C2=54\\ C3=51\end{tabular}  & \multicolumn{1}{l|}{\begin{tabular}[c]{@{}l@{}}C1=18,C2=10,\\ C3=21,C4=7,\\ C5=34,C6=18,\\ C7=12,C8=17,\\ C9=6,C10=16\end{tabular}} & \begin{tabular}[c]{@{}l@{}}C1=38,C2= 42,\\  C3,79\end{tabular} & \multicolumn{1}{l|}{\begin{tabular}[c]{@{}l@{}}C1=44,C2=2,\\ C3=21,C4=31,\\ C5=58, C6=1,\\  C7=2\end{tabular}} & \begin{tabular}[c]{@{}l@{}}C1=20,C2=19,\\ C3=13,C4=29,\\ C5=19,C6=41,\\ C7=18\end{tabular} & \begin{tabular}[c]{@{}l@{}}C1=41,C2=45,\\ C3=63,C4=3,\\ C5=7\end{tabular}      \\
\multicolumn{1}{l|}{Cluster summary (mode method)}              & \begin{tabular}[c]{@{}l@{}}C1=57,C2=60,\\ C3=42\end{tabular} & \multicolumn{1}{l|}{\begin{tabular}[c]{@{}l@{}}C1=44,C2=3,\\ C3=22,C4=51, \\ C5=11,C6=23,\\ C7=5\end{tabular}}                      & \begin{tabular}[c]{@{}l@{}}C1=36,C2=34,\\ C3=89\end{tabular}   & \multicolumn{1}{l|}{\begin{tabular}[c]{@{}l@{}}C1=49 ,C2= 17,\\ C3=32 , C4=61\end{tabular}}                            & \begin{tabular}[c]{@{}l@{}}C1=25,C2= 51,\\  C3=11,C4= 65,\\ C5= 7\end{tabular}             & \begin{tabular}[c]{@{}l@{}}C1=1,C2=42,\\ C3=47,C4=63,\\ C5=2,C6=4\end{tabular} \\ \hline
\end{tabular}}
\end{table}

\section{Real Data Analysis} \label{real}
In this section, we provide a comprehensive overview of our findings derived from applying the BGWR model in a real-world scenario. We explain the sources of the data employed and describe the cluster configurations associated with the BGWR parameter coefficients. Moreover, we present a thorough investigation into preschool attendance as estimated by the BGWR model, alongside the corresponding probability values. We also discuss the incorporation of dynamic variable selection into our analysis of the actual data, and we present a visual representation of substantively important coefficients for each region.
\subsection{Sources of the data }
The Children’s Health Queensland (CHQ) has created an impressive resource called the CHQ Population Health Dashboard, which provides data on key health outcomes and socio-demographic factors for a one-year period from 2018-2019. The dashboard is based on information from 528 small areas (statistical area level SA2) across the state of Queensland, Australia. The dashboard includes over 40 variables, visualized in a user-friendly format. The case study considered in this paper focuses on the health outcomes section of the dashboard, specifically, vulnerability indicators, which measure developmental vulnerability for children across five Australian Early Development Census (AEDC) domains: 
\begin{itemize}[topsep=1pt]
\item Physical health and wellbeing, which evaluates children's physical readiness for school, their level of physical independence, and their gross and fine motor skills.
\item Social competence, which assesses children's overall social competence, responsibility, respect, approaches to learning and readiness to explore new things.
\item Emotional maturity, which examines children's pro-social and helping behaviour, anxious and fearful behaviour, aggressive behaviour, and hyperactivity and inattention.
\item Language and cognitive skills (school-based), which evaluates children's basic literacy, interest in literacy, numeracy, and memory, as well as their advanced literacy and basic numeracy.
\item Communication skills and general knowledge, which measures children's communication skills and general knowledge.
\end{itemize}
\noindent The AEDC data also includes two additional indicators: vulnerable on one or more domains (Vuln 1) and vulnerable on two or more domains (Vuln 2). Additionally, the CHQ dashboard encompasses data on socio-demographic factors that may be linked to health outcomes. Three factors considered in the analysis are the Socio-Economic Indexes for Areas (SEIFA) score, attendance at a preschool, and remoteness. The SEIFA score is a socio-economic index that summarizes a variety of data on individual and family economic and social conditions in a given area. It ranges from 1 to 5, with a low score indicating that greater disadvantage. The remoteness factor includes the categories of cities,  regional, and remote. In 2018, Queensland had 294 SA2s in cities, 208  regional SA2s, and 24 remote SA2s. The analysis uses data from the AEDC, which is conducted every three years and collects data on children in their first year of school. The data used in this study is from the 2018 census. Due to the aggregated nature of the data, the analysis focuses on the proportion of vulnerable children within each SA2 \cite{snapshot}, collected by first-year teachers across the Australian government and non-government schools, with parents' agreement. 
The final dataset therefore compromises the proportion of children who attended preschool, the SEIFA, and remoteness for each SA2.

Between 3 and 6\% of the data had missing variables, which were imputed using the average of the proportions from the neighbouring SA2s. For categorical data, such as remoteness, missing values were imputed using the highest frequency category of the neighbouring SA2s. However, missing values for two islands could not be imputed as these regions have no contiguous neighbours. As a result, the analysis was carried out on the remaining 526 SA2 areas.
\\ \textbf{Ethical considerations}
The study was approved by the Human Research Ethics Committee of Children’s Health Queensland Hospital and Health Service (HREC/21/QCHQ/75725).
\subsection{Case study analysis}
The main objective of the analysis was to examine the factors that influence children's developmental vulnerability in one or more domains (Vuln1) with a particular focus on the importance of attendance at preschool. Using the proposed approach, we aimed to identify clusters of regions that are similar with respect to the influence of attendance at preschool on Vuln1.

The previous work on the BGWR model was designed to deal with continuous covariates \cite{ma2021geographically}. However, in this paper, we extended the BGWR model to deal with categorical variables as well, specifically for the "remoteness factor". Since a region can only belong to one of the three levels (cities, regional, or remote), we created a list of covariates for each region for inclusion in the BGWR model. This avoids the critical issue of drawing from the prior when there is no valid information available for a particular category.

In Section \ref{inf}, we explore inferences associated with the proportion of attendance at preschool, as derived from the BGWR model. In section \ref{RJMCMC}, we explore the efficiency of our variable selection algorithm. Finally, in Section \ref{clust}, we discuss the probabilistic clustering results from the two algorithms. Additionally, we investigate the probability of each region belonging to specific clusters.
\subsubsection{Inferences from the Bayesian geographically weighted regression} \label{inf}
 We use the graph distance as well as the greater circle distance in the BGWR model. So, we adopted a non-informative prior for the bandwidth and estimated the optimal bandwidth as part of the analysis by choosing a value of $D=100$ and the maximum distance between any two points is 10. 
 
\begin{figure} [ht!]
    \centering
    \includegraphics[scale=0.32]{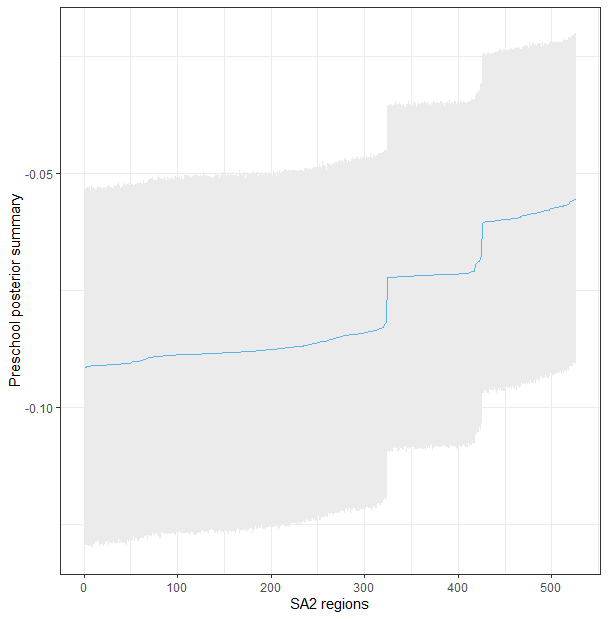}
    \includegraphics[scale=0.25]{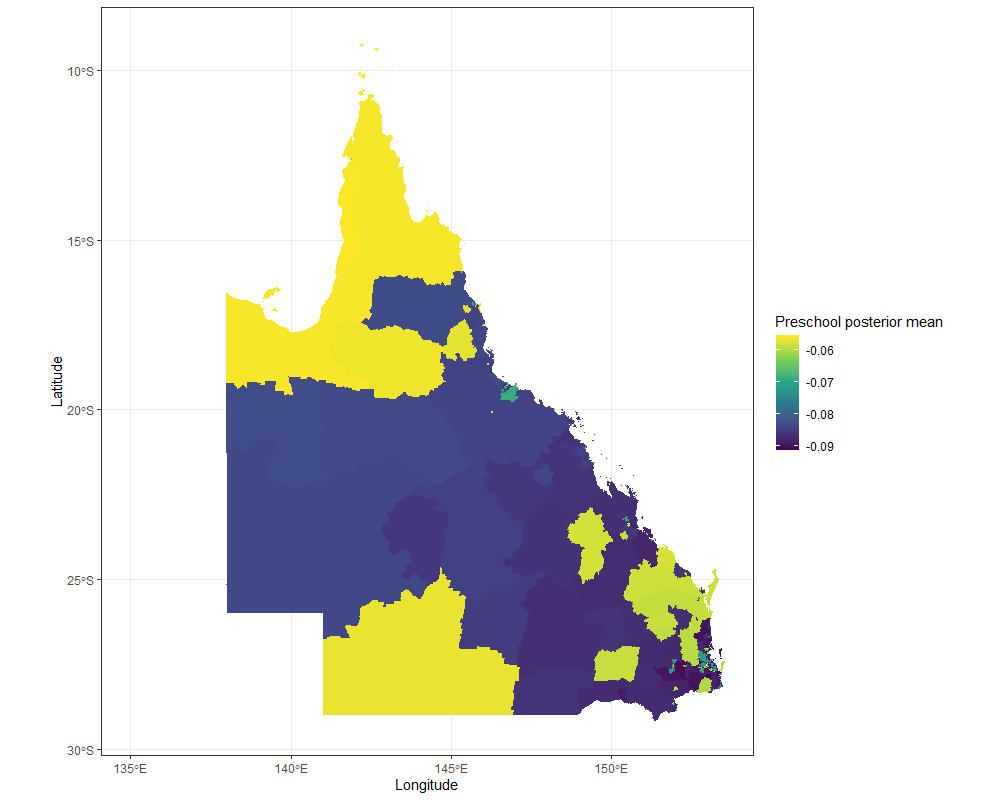}
     \caption{LHS) posterior means and 95\% credible interval for the coefficients associated with preschool attendance, RHS) the geographical distribution of the posterior means.}
    \label{fig:my_label_1}
\end{figure}

 The MCMC algorithm was run for 400,000 iterations with a burn-in period of 20,000 iterations. The results reported are based on the remaining 380,000 iterations. We focused on exemplar insights from the BGWR model. First, the posterior mean and 95\% credible interval were obtained for the coefficients associated with attendance at preschool in each region. The results are summarised in Figure \ref{fig:my_label_1}. 
It appears that there is a substantive negative relationship between the proportion of attendance at preschool and Vuln 1. This negative association is stronger in the South-East of Queensland, particularly in greater Brisbane, and comparatively weaker in the northern regions of Queensland. Several factors could contribute to this relationship, such as parental background, Indigenous status, and access to preschool services. The second insight is on the evaluation of the non-stationary variance.

\begin{figure}[ht!]
    \centering
    \includegraphics[scale=0.4]{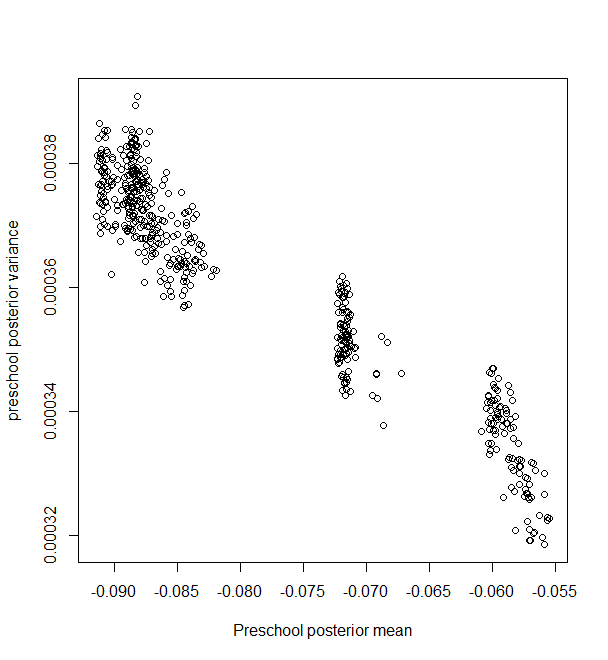}
    \caption{The scatter plot illustrates the relationship between the mean and variance per each region, derived from the posterior of the coefficients associated with attendance at preschool in the BGWR model. }
    \label{fig:my_label30}
\end{figure}

\begin{figure}[ht!]
    \centering
    \includegraphics[scale=0.5]{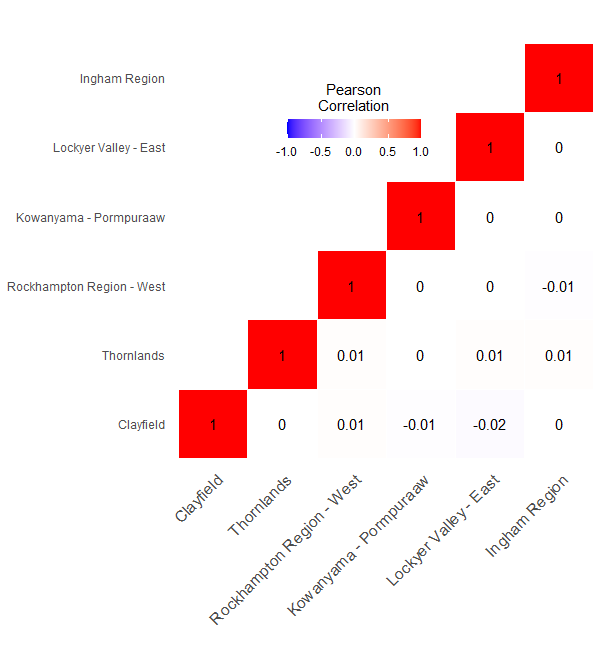}
    \caption{The correlation between the posterior distributions of the attendance at preschool parameter across the six regions.}
    \label{fig:my_label2}
\end{figure}

\begin{figure} [ht!]
    \centering
    \includegraphics[scale=0.4]{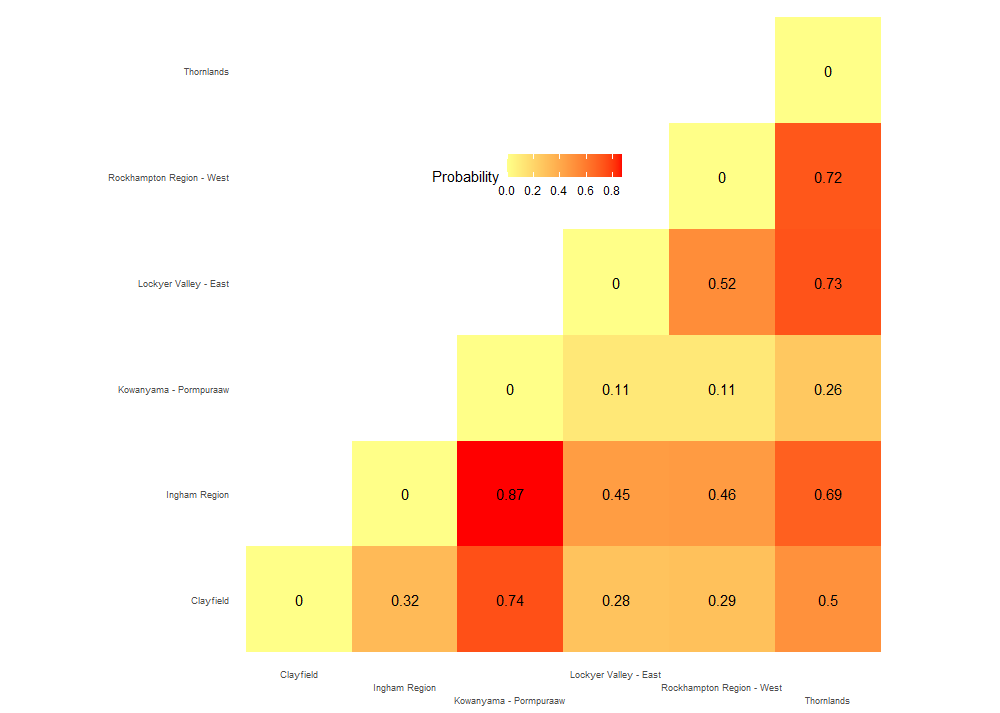}
    \caption{Comparison of probabilities for attendance at preschool posterior across the selected six regions.}
    \label{fig:my_label3}
\end{figure}

Figure \ref{fig:my_label30} presents a scatter plot of the posterior means and variances per region derived from the BGWR model. Regions with smaller posterior means generally have larger variances. Noticeably, this relationship is affected by the three levels of remoteness factor in the case study. 
The third insight involves comparisons between various regions. We selected two regions from cities, two outer regions, and two remote areas, for the purpose of identifying differences and similarities. These regions include Thorlands, Clayfield, Rockhampton region - West, Lockyer Valley-East, Ingham region, and Kowanyama Pormpuraaw. A visualization of the correlation between these six regions is shown in  Figure \ref{fig:my_label2},  which reveals a slight association among the posterior estimates for preschool attendance. Figure \ref{fig:my_label3} presents a comparative analysis of probabilities among the six regions regarding the posterior estimates of the coefficient associated with attendance at preschool. 
The probabilities are computed by comparing the posterior draws of the coefficient for one region with those of other regions, for each MCMC simulation.
The vulnerabilities in our study are expressed as proportions ranging between 0 and 1. Given this scale, large coefficient values would not only be unexpected but also potentially misleading. Instead, smaller coefficients are more appropriate and consistent with the data's inherent structure.
Additionally, while the range of these coefficients might appear limited, even small variations can be meaningful. Especially in a context where the dependent variable operates on such a narrow scale, even slight changes in predictor values can lead to significant shifts in outcomes. Therefore, allowing the coefficients to vary, even within a small range, is essential. This flexibility ensures that our model can capture and represent subtle spatial patterns and gradients in vulnerabilities, which might be critical for understanding and addressing them effectively.

\begin{figure} [ht!]
    \centering
    \includegraphics[scale=0.4]{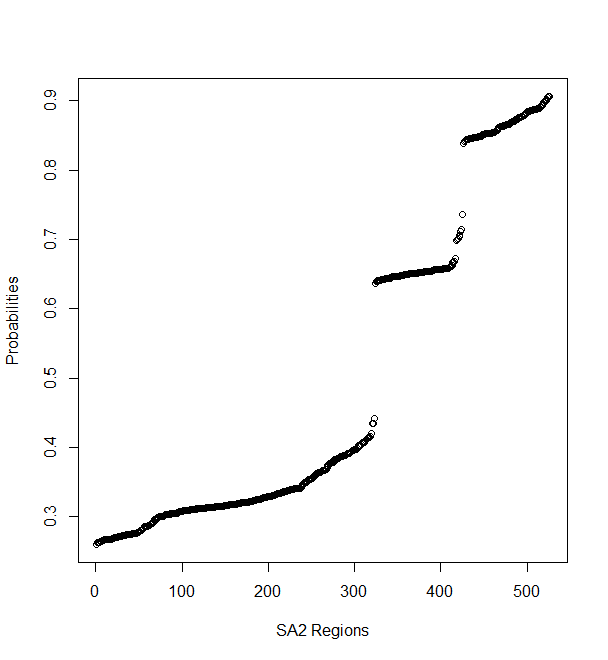}
    \caption{ The probabilities of each region to get the posterior draws coefficient associated with preschool attendance greater than its posterior mean.}
    \label{fig:my_label4}
\end{figure}

Assume $\beta_p(i)$ is the coefficient associated with attendace at preschool and ($s$) location. In Figure \ref{fig:my_label3}, the values are computed using the formula \( P_r(\beta_p(k) > \beta_p(q)) \). This formula is computed over \( M\) iterations when the $\beta_p$ coefficient of the region \( k \) is greater than the coefficient in region \( q \). Specifically, for each iteration, an indicator function assigns a value of 1 when \( \beta_p(k) > \beta_p(q) \) and 0 otherwise. The final probability is the sum of these indicator values divided by the total number of iterations, \( M \). For example, the value at row 5, column 3 (0.87) represents the probability of attendance at preschool in Kowanyama Pormpuraaw to be greater than Ingham region.\\
The analysis was extended to explore $P_r(\beta_p(i)> \bar{\beta_p})$, where $\beta_p(i)$ is similar to before, and $\bar{\beta_p}$ denotes the attendance at the preschool posterior mean for specific region $i$. The resulting probabilities are visualized in Figure \ref{fig:my_label4}, offering insights into the variability and disparities among regions. Over 300 SA2 regions have a probability of less than 0.5 exceeding their posterior mean $\bar{\beta_p}$.

\begin{figure} [ht!]
    \centering
    \includegraphics[scale=0.5]{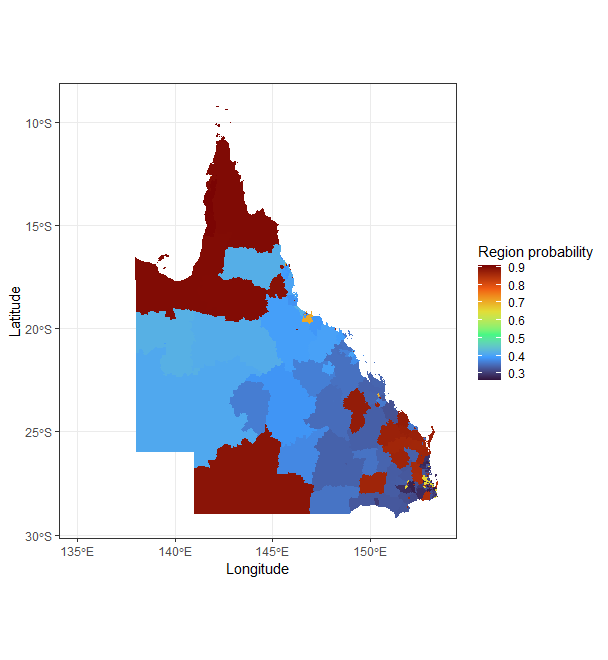}    \caption{The spatial distribution of the probability of each region to get the posterior draws coefficient associated with preschool attendance greater than its posterior mean.}
    \label{fig:my_label5}
\end{figure}

The distribution of these probabilities is depicted in Figure \ref{fig:my_label5}. Regions with higher probabilities are highlighted, indicating a greater probability of exceeding the posterior mean for that particular region.
In this plot, 101 SA2 regions exhibit a probability greater than 0.8 of having $(\beta_p(i)> \bar{\beta_p})$  across $M$ iterations.

\begin{figure} [ht!]
    \centering
    \includegraphics[scale=0.6]{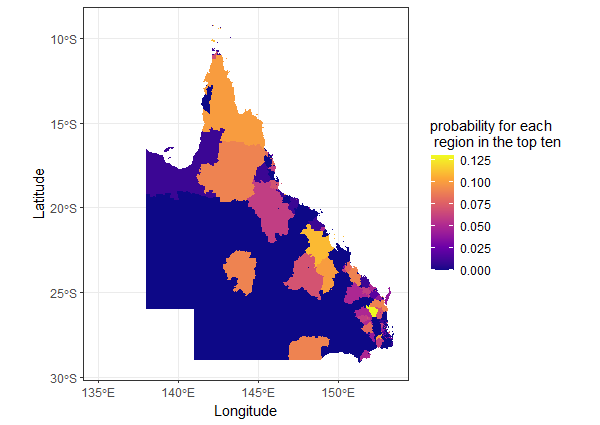}
    \caption{The probability of each region on the map to be in the top ten.}
    \label{fig:my_label6}
\end{figure}

Additionally, our analysis involved calculating the probability of each region being among the top ten regions for the coefficients associated with attendance at preschool $\beta_p$. To achieve this, we performed the following steps: For each iteration in our study (denoted by $M$), we obtained a posterior draw for $\beta_p$ in each of the 526 SA2 regions. These draws were then sorted, and the top ten were identified. An indicator variable was created for each iteration $M$, taking the value 1 if it was among the top ten and 0 otherwise. This process was repeated for all iterations in the posterior. After obtaining the indicator values for all iterations, we calculated the probabilities by summing up the values corresponding to a specific region and dividing by the total number of iterations. This gave us the probabilities of each region being in the top ten. The probability of each region being in the top ten is shown in Figure \ref{fig:my_label6}. This visual representation highlights that the far north of Queensland stands has a stronger probability of being in the top ten compared to South-East Queensland.

Finally, an assessment was undertaken of the impact of different kernel functions in the BGWR analysis of the case study. The exponential, bi-square and Gaussian functions were considered. Results were presented in Table \ref{tab:my-table5}. As in the simulated study in section \ref{simulated}, Table \ref{tab:my-table3}, the performance of the algorithm is evaluated based on three criteria: WAIC, DIC  and $P_D$ (effective sample size). The table shows that the exponential kernel outperforms the other two kernels, bi-square and Gaussian, in all three criteria.  Furthermore, we also observed a consistent pattern in the behaviour of the greater circle distance and the graph distance when utilised with non-informative priors for the bandwidth in both our case study and simulated data. The results from both distance metrics provide similar posterior estimates.

\begin{table}[ht!] 
\caption{\label{tab:my-table5} WAIC, DIC, and effective sample size values from the proposed BGWR algorithm using different kernels for the case study.}
\centering
\begin{tabular}{llll}
\hline
           & Exponential kernel & Bi-square kernel & Gaussian kernel  \\ 
\hline
WAIC       & -801304.5          & -802014.9        & -801161.4 \\
DIC        & -742933.6          & -743896.3        & -744056.8 \\
\(P_D\)   & 3253.3             & 3176.5           & 3176.5 \\
\hline
\end{tabular}
\end{table}

\subsubsection{Dynamic variable selection: real data analysis} \label{RJMCMC}
\begin{figure}[ht!]
    \centering
   \includegraphics[scale=0.5]{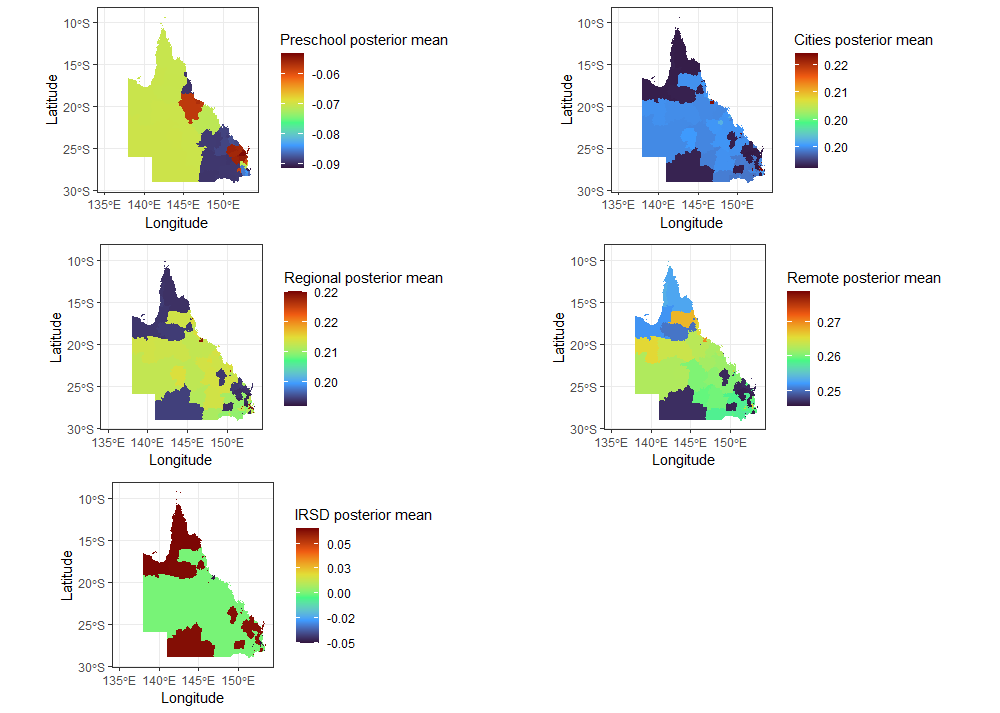}
    \caption{
    The posterior summary from the proposed selection algorithm.}
    \label{fig:enter-labelww}
\end{figure}

\begin{figure}[ht!]
    \centering
    
     \includegraphics[scale=0.5]{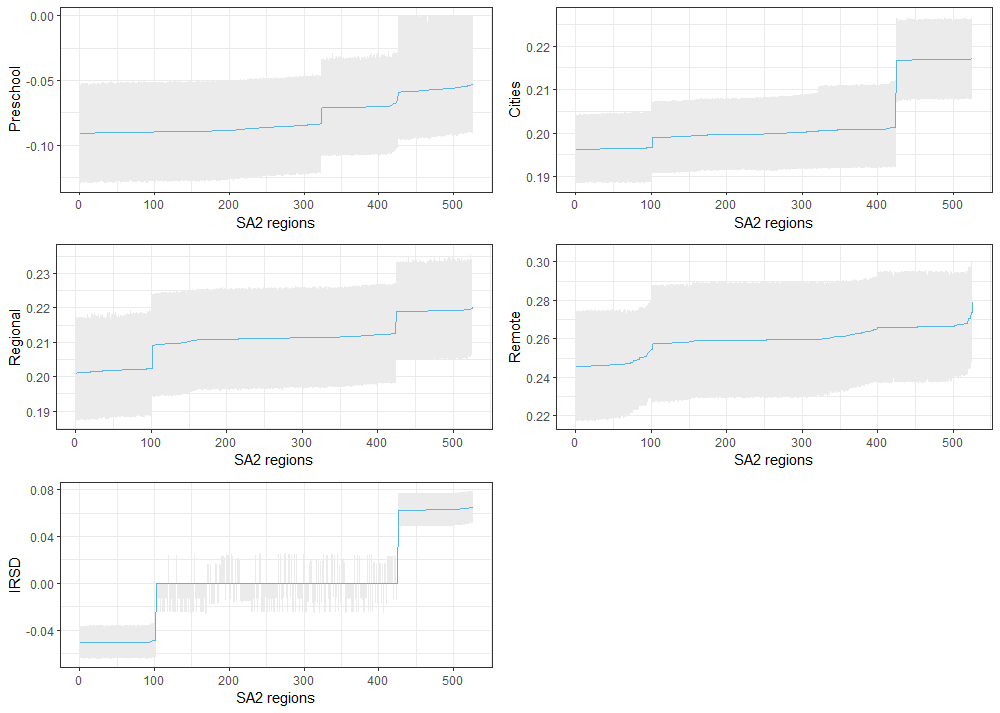}
    \caption{The posterior mean summary for each covariate from the local model according to the proposed selection algorithm.}
    \label{fig:enter-label22}
\end{figure}

The algorithm described in Section \ref{algorithm2} was applied to the real case study. The summarized results presented in Figure \ref{fig:enter-label22} show that the preschool factors have a substantive negative impact, particularly in South-East and Central Queensland. In addition, the IRSD covariate is also important for some specific locations in the case study. Its effect becomes more obvious when focusing on southeast Queensland, and the far north of Queensland. A summary of the posterior estimates obtained from the model is presented in Figure \ref{fig:enter-labelww}.

\subsubsection{Probabilistic cluster analysis and its insights}
\label{clust}
 We extracted 500 random iterations from the posterior distributions obtained from the BGWR analysis and applied GMM and DPMM clustering algorithms for all the regression coefficients at each iteration. To determine the optimal number of clusters for the GMM, we employed the BIC across these samples. The analysis revealed that the optimal number of clusters is 4, and we present the optimal number of clusters in Figure \ref{fig:enter-label5} from Appendix \ref{real_further_test}.

\begin{figure}[ht!]
    \centering
    \includegraphics[scale=0.5]{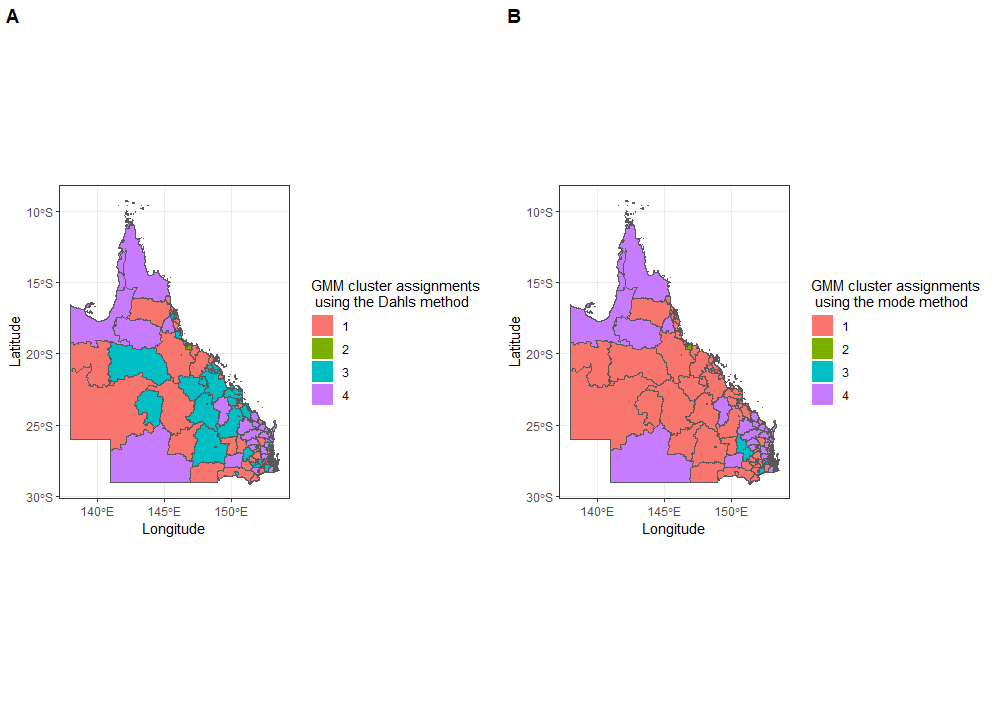}
    \caption{Spatial pattern of GMM clusters: case study results using Dahl's and mode methods.}
    \label{fig:enter-labelqqq}
\end{figure}

Using  Dahl's method, we obtained four clusters with sizes, 103, 101, 220, and 102, whereas, using the mode we also obtained four clusters with different sizes 303, 102, 20, and 101. In our case study, Dahl's method yielded more evenly sized clusters. The mode method, on the other hand, resulted in one notably large cluster and one exceptionally small one, making Dahl's method the more consistent of the two. The spatial distribution of these cluster assignments using Dahl's method and the mode can be observed in Figure \ref{fig:enter-labelqqq}.

\begin{figure} [ht!]
    \centering
    \includegraphics[scale=0.4]{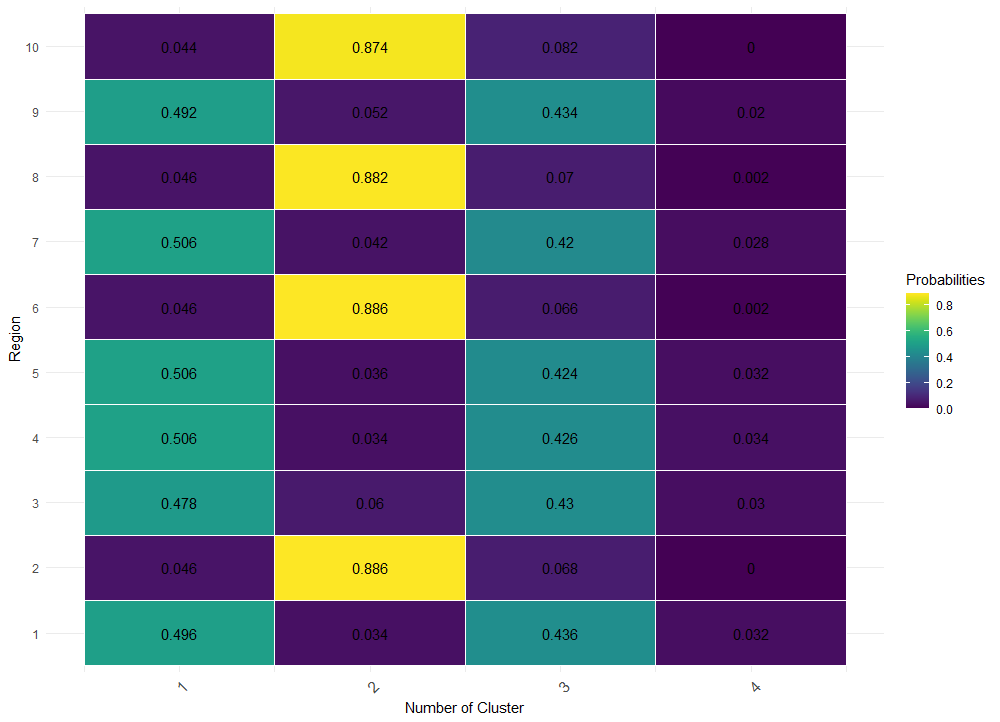}
    \caption{The probabilities of each region in Queensland belonging to specific clusters obtained from the GMM.}
    \label{fig:enter-labelqqqqq}
\end{figure}

We present the probability of each region belonging to a specific cluster from the GMM compared with the mode method in the heatmap. Figure \ref{fig:enter-labelqqqqq} shows these probabilities for ten regions in Queensland. Each row in the heatmap represents one of the four possible values (1, 2, 3, 4) indicating the clusters, while each column corresponds to one of the first ten regions in Queensland. The colour intensity in each cell represents the probability of the corresponding region belonging to the respective cluster: lighter shades signify higher probabilities, while darker shades indicate lower probabilities. A complete view of the probabilities for all regions in Queensland is provided in Appendix \ref{real_further_test}. 

\begin{figure}[ht!]
    \centering
    \includegraphics[scale=0.4]{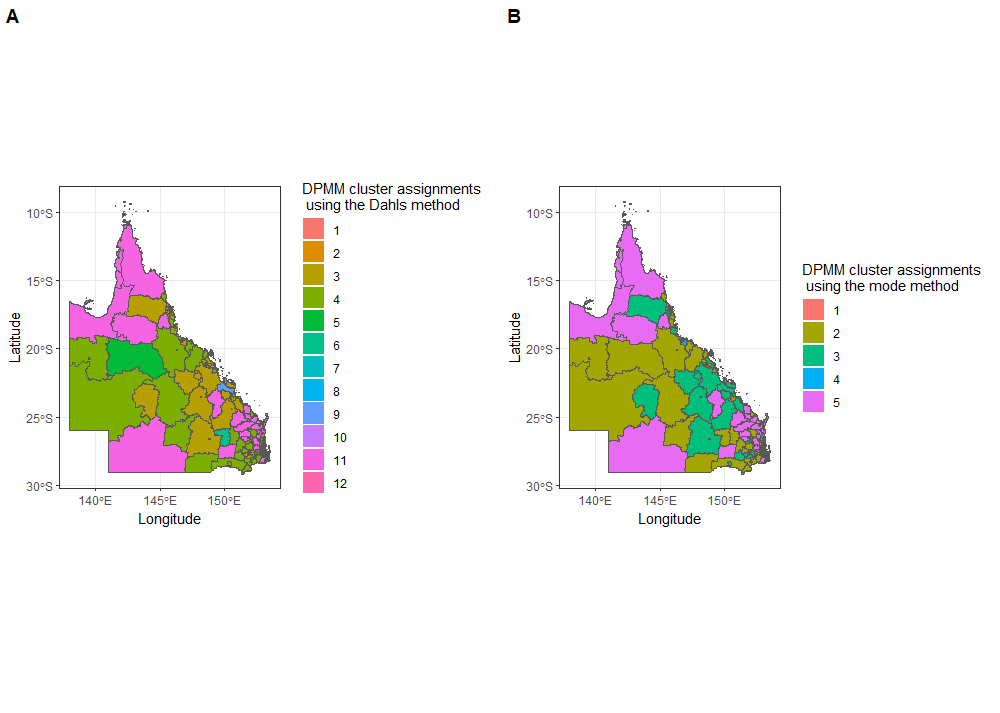}
    \caption{The spatial distribution of the cluster configuration obtained from DPMM using Dahl's method as well as the mode method.}
    \label{fig:enter-labelqq}
\end{figure}

\begin{figure}[ht!]
    \centering
    \includegraphics[scale=0.4]{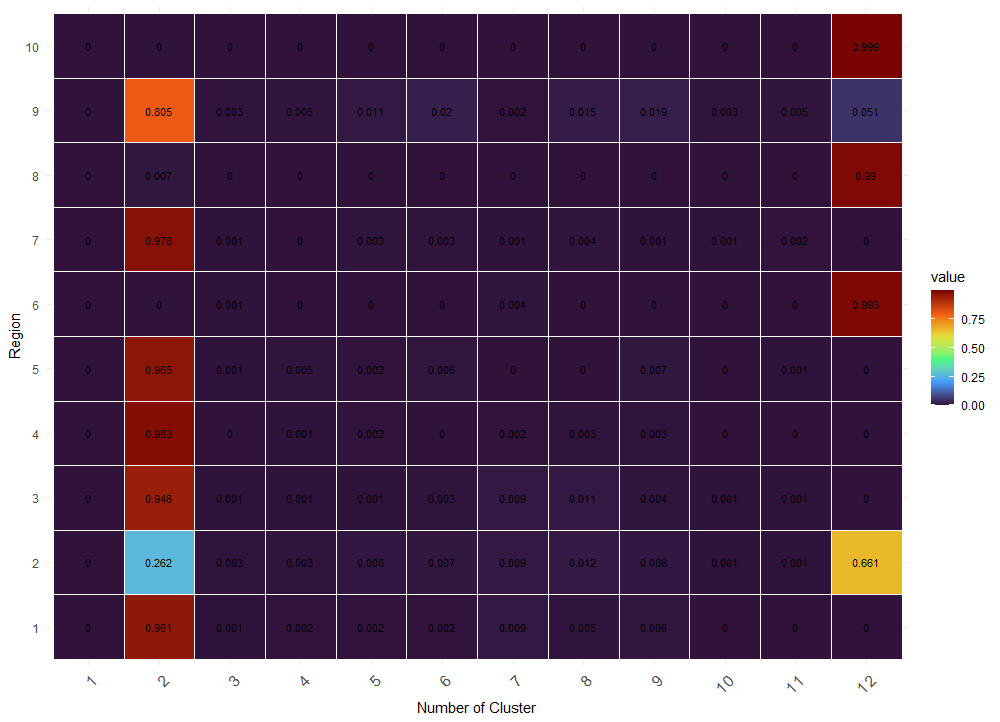}
    \caption{  The probabilities of each region in Queensland belonging to specific clusters obtained from the DPMM.}
    \label{fig:enter-labele}
\end{figure}

The cluster configuration process of DPMM happens in two stages: The first stage is a within-sample cluster configuration, obtained for each of the 500 randomly chosen iterations in the MCMC analysis of the DPMM. The configurations are based on Dahl's method and the mode method. The second step is across-sample cluster configuration. In this stage, we consider the entire set of 500 samples and their corresponding cluster configurations obtained from the first stage and perform clustering configurations again to get the final cluster configuration for each region (see Appendix \ref{DPMMapp}). 
For our case study, the final cluster number using Dahl's method is 12 with sizes as follows: 11, 196, 1, 1, 2, 1, 1, 107, 3, 3, 100, and 100. The mode method also found five clusters with sizes 1, 212, 109, 101, and 103. Dahl's method resulted in more clusters. However, the maps from both methods show overlapping regions, with specific differences highlighted in Figure \ref{fig:enter-labelqq}. 
We also calculated the probabilities of each region belonging to one of the 12 clusters using Dahl's method. Figure \ref{fig:enter-labele} presents the probabilities for the first ten regions. The complete probabilities are available in Appendix \ref{real_further_test}. 
\section{Discussion}
We have explored a clustered Bayesian GWR approach for estimating  and clustering regression coefficients, in the context of spatially heterogeneous data. The model accommodates both continuous and categorical variables. The new vectorisation provides remarkable improvements in MCMC sampling efficiency. This enables scalable analysis of large-scale datasets without requiring the dataset to be split into smaller sub-regions. An additional feature of the BGWR model is the facility to identify  covariates that are important in some locations but have minimal impact in other locations. \\
The extension of the approach to include probabilistic clustering  substantially expands the utility of the analysis and provides insights about similar geographic regions that are  of great interest in subdomains such as health.

 The proposed methods were successfully implemented in R using the nimble computational framework \cite{de2017programming}. Additionally, we investigated various weighting schemes based on graph distance and greater circle distance, finding that these approaches yield models with robust parameter estimation capabilities, particularly when facing spatially heterogeneous data.

There are several avenues for future research that extend beyond the scope of this paper. For instance,  our Bayesian approach could be extended to handle generalized linear models (GLMs), which would significantly enhance its applicability to a broader range of regression models and diverse data types. This extension would involve adapting the algorithm to accommodate various link functions and likelihood distributions \cite{kowal2022bayesian}. Additionally, integrating penalized methods like ridge regression, lasso, or elastic net regularization into our approach could efficiently handle high-dimensional data and potentially improve the accuracy and stability of parameter estimation.  Another direction is to develop a robust framework to determine optimal bandwidths.  A comprehensive study on bandwidth selection is essential to ensure the efficiency and accuracy of the BGWR algorithm in handling spatially varying covariate effects within a Bayesian framework \cite{da2018comparing}.

For model interpretability and identifying relevant covariates, appropriate approaches for variable selection in the context of clustered regression should be explored. Utilising DPMM and GMM to obtain clustering information of regression coefficients shows promise but requires further investigation to resolve inconsistencies in the posterior on the number of clusters \cite{geng2019probabilistic}. Moreover, to enhance the reliability and robustness of the proposed algorithm, especially when dealing with spatially dependent data and clustered structures, a prior for spatially clustered regression coefficients should be developed \cite{kowal2022bayesian,geng2019probabilistic}. 

Developing a comprehensive spatial probabilistic framework would provide a solid foundation for handling complex spatial structures, incorporating various spatial components, and enabling integration with other spatial models and techniques. For example, investigating the connection of the proposed BGWR with latent variable models might lead to more advanced and efficient clustering techniques. Such integration could allow for more flexible and adaptive clustering approaches that account for complex spatial patterns \cite{duan2007generalized}.
Finally, another promising direction to extend this work is to develop the proposed algorithm for a multivariate BGWR model. By considering multiple response variables simultaneously, we can enhance the model's capability to capture complex spatial dependencies and better understand the relationships between the variables in a spatial context. This endeavour holds great promise for advancing the state-of-the-art in spatial statistics and providing a powerful tool for tackling real-world spatial data challenges.
\section{Conclusion}
In this paper, we have achieved three main contributions. The first is the combination of Bayesian Geographically weighted regression with clustering via the Gaussian mixture model and the Dirichlet process mixture model to detect patterns in how different factors influence geographic data. The second is the optimisation of the computational algorithm to work well with large geographic areas and many spatial regions, leading to faster estimations. The third is the inclusion of the dynamic variable selection for each location in the BGWR model.

In addition to validating the effectiveness of our method through substantive simulated studies, we considered a case study focusing on children's development in Queensland. Real data from Children's Health Queensland and Australian Early Development Census were utilised for this study. By examining vulnerability in at least one of five critical development domains—physical health and wellbeing, social competence, emotional maturity, language and cognitive skills (school-based), and communication skills and general knowledge—our approach successfully identified clusters of regions with similar developmental vulnerabilities. This discovery holds significant implications for health services planning and early intervention efforts, aiming to enhance the overall wellbeing and success of children during their formative years. Furthermore, we conducted an in-depth exploration of the preschool attendance covariate using the BGWR model.

Overall, our research contributes to the field of spatial regression analysis by offering a powerful and efficient tool for understanding spatially varying patterns in the relationship between risk variables and responses and opens up new avenues for exploring spatial clusters and their implications. Our approach can be used in  research in various domains beyond child development, such as health services planning and spatial analysis in general.

\newpage
\begin{appendices}
\section{Spatial Distances} \label{distances}
When working with areal data, the graph distance is an alternative distance metric that can be used. It is based on the concept of a graph, where $V=\{v_1,...,v_m\}$ represents the set of nodes (vertices) and $E(G)=\{e_1,...,e_n\}$ represents the set of edges connecting these nodes. The graph distance is defined as the distance between any two nodes in the graph,

\[ d_{v_sv_i}=\begin{cases} 
      |V(e)| &   \text{if $e$ is the shortest distance connecting a pair of nodes.}\\
      \infty & \text{if the two nodes are not connected.}
   \end{cases}
\]
where $|V(e)|$ is the number of edges in $e$  \cite{gao2010survey}. Choosing the right settings for distance methods can be tricky and depend  on "how near is really near" One way to decide this is by using graph distance. If regions have common boundaries, they have a graph distance of 1, which means they are close. But if they have a distance of more than 1, they are not so close. Regions that aren't very close should get less importance when we do calculations. The graph distance-based weighted function is given:
\[ W(s)=\begin{cases} 
      1 &   \text{if $d_i(s)\leq b$}\\
      f(d_i(s),b) & \text{ otherwise}
   \end{cases}
\]
where $d$ is the graph distance, $f$ is a weighting function, and $b$ represents the bandwidth. In this study, we suppose that $f()$ is a negative exponential function \cite{ma2021geographically}, so that:
\[ W(s)=\begin{cases} 
      1 &   \text{if $d_i(s) \leq 1$}\\
      e^{(-d_i(s)/b)} & \text{ otherwise}
   \end{cases}
\]
 
Another way to calculate the distance from the areal data is the greater circle distance (GCD) which calculates the shortest distance $d_i(s)$ between two points on the surface of a sphere (e.g., Earth) using their latitude and longitude coordinates. This accounts for the curvature of the Earth and gives the distance along the surface of the sphere, following the path of a great circle. The great circle represents the largest circle that can be drawn on a sphere and passes through the two points. The GCD is given as \cite{carter2002great} :
\begin{equation}
d_i(s) = r \cdot \cos^{-1} \left( \cos(a_s) \cdot \cos(a_i) \cdot \cos(b_s - b_i) + \sin(a_s) \cdot \sin(a_i) \right)
\label{GC}
\end{equation}

To calculate the GCD, the formula involves trigonometric functions. Specifically, it calculates the arc-cosine of a term involving the cosines and sines of the latitudes and longitudes of the two points. 
$a_s$ and $b_s$ represent the latitude and longitude for location $s$, and $a_i$ and $b_i$ represent the latitude and longitude for location $i$. The term inside the arc-cosine represents the dot product of two vectors in a 3D space, which is used to determine the angle between the two points. The inverse cosine function of this dot product yields the angle, and by multiplying it with the Earth's radius $r$, we get the great circle distance between the two points \cite {carter2002great}.\\
The weighting scheme introduced in graph distance and greater circle distance, where neighbours are assigned the same weight and all others receive positive weight, provides additional assurance of parameter estimation stability when compared to the other weighting schemes.
\section{Metropolis-Hastings Sampling for Bayesian Geographically Weighted Regression (BGWR)}    \label{real_nimble}
This section explains how to draw samples from the posterior distribution of the BGWR model parameters, considering the spatial dependencies, likelihood function, and prior distributions.\\
    \textbf{Initialization:} 
   \begin{itemize}
       \item Start with an initial set of values for all parameters: $\beta(s)$, $\sigma^2(s)$, $\sigma_\beta$, and $b$.
   \end{itemize}
   
 \textbf{Iterative Sampling Process:} For each iteration until the desired number of samples:
 
    \begin{itemize}
    \item \textbf{Proposal Generation:} Propose new values for each model parameter based on their respective proposal distributions and a random walk step:
    \begin{itemize}
        \item Coefficients \( \beta(s) \): propose \( \beta'(s) = \beta(s) + \epsilon \) where \( \epsilon \sim N(0, \sigma_\beta^2) \).
        \item Spatial Variance \( \sigma^2(s) \): given the prior is \( \sigma^2(s) \sim \text{IGamma}(\alpha_1, \alpha_2) \), propose \( \sigma'^2(s) = \sigma^2(s) + \delta \) where \( \delta \) is drawn from some distribution (e.g., Gaussian).
        \item Coefficient Variance \( \sigma_\beta \): given the prior is \( \sigma_\beta \sim \text{IGamma}(\alpha, \beta) \), propose \( \sigma_\beta' = \sigma_\beta + \zeta \) where \( \zeta \) is drawn from some distribution.
        \item Bandwidth \( b \): propose \( b' = b + \eta \) where \( \eta \) is drawn from some distribution (e.g., uniform).
    \end{itemize}
\end{itemize}
        
     \textbf{Compute Acceptance Ratio $R$:}

\begin{itemize}

    \item \textbf{Coefficients} \( \beta(s) \):
        \begin{align*}
            R_\beta &= \frac{\pi^{-n/2} \cdot \sigma'^{-n}(s) \cdot |W(s)|^{-1/2} \exp\left(-\frac{1}{2} \sum_{i} w_i^{-1}(s) \cdot (y(s) - x^T(s)\beta'(s))^2\right) \times P(\beta'(s))}{\pi^{-n/2} \cdot \sigma^{-n}(s) \cdot |W(s)|^{-1/2} \exp\left(-\frac{1}{2} \sum_{i} w_i^{-1}(s) \cdot (y(s) - x^T(s)\beta(s))^2\right) \times P(\beta(s))}
        \end{align*}

        \item \textbf{Compute Acceptance Ratio} \( R_{\sigma^2} \):
        \begin{align*}
            R_{\sigma^2} &= \frac{\pi^{-n/2} \cdot \sigma'^{-n}(s) \cdot |W(s)|^{-1/2} \exp\left(-\frac{1}{2} \sum_{i} w_i^{-1}(s) \cdot (y(s) - x^T(s)\beta(s))^2\right) \times P(\sigma'^2(s))}{\pi^{-n/2} \cdot \sigma^{-n}(s) \cdot |W(s)|^{-1/2} \exp\left(-\frac{1}{2} \sum_{i} w_i^{-1}(s) \cdot (y(s) - x^T(s)\beta(s))^2\right) \times P(\sigma^2(s))}
        \end{align*}

        \item \textbf{Compute Acceptance Ratio} \( R_{\sigma_\beta} \):
        \begin{align*}
            R_{\sigma_\beta} &= \frac{\pi^{-n/2} \cdot \sigma'^{-n}(s) \cdot |W(s)|^{-1/2} \exp\left(-\frac{1}{2} \sum_{i} w_i^{-1}(s) \cdot (y(s) - x^T(s)\beta(s))^2\right) \times P(\sigma'_\beta)}{\pi^{-n/2} \cdot \sigma^{-n}(s) \cdot |W(s)|^{-1/2} \exp\left(-\frac{1}{2} \sum_{i} w_i^{-1}(s) \cdot (y(s) - x^T(s)\beta(s))^2\right) \times P(\sigma_\beta)}
        \end{align*}

        \item \textbf{Compute Acceptance Ratio} \( R_b \):
        \begin{align*}
            R_b &= \frac{\pi^{-n/2} \cdot \sigma^{-n}(s) \cdot |W(s)|^{-1/2} \exp\left(-\frac{1}{2} \sum_{i} w_i^{-1}(s) \cdot (y(s) - x^T(s)\beta(s))^2\right) \times P(b')}{\pi^{-n/2} \cdot \sigma^{-n}(s) \cdot |W(s)|^{-1/2} \exp\left(-\frac{1}{2} \sum_{i} w_i^{-1}(s) \cdot (y(s) - x^T(s)\beta(s))^2\right) \times P(b)}
        \end{align*}

\end{itemize}

     \textbf{Evaluate} 
     \begin{itemize}
         \item  \( R \) (be it \( R_\beta \), \( R_{\sigma^2} \), \( R_{\sigma_\beta} \), or \( R_b \)).
  \end{itemize}
     \textbf{Accept or Reject} 
     \begin{itemize}
         \item Draw a random value $u$ from a uniform distribution $U(0,1)$. If $u < \min(1, R)$, accept the proposed parameters. Otherwise, retain the current parameter values.
   \end{itemize}

    \textbf{Diagnostics} 
    \begin{itemize}
        \item Repeat the above steps for the desired number of iterations until converge.
    \end{itemize}

\section{RJMCMC Algorithm}    \label{RJMCMCapp}
RJMCMC lets the chain switch between models with different parameter counts. In this paper, RJMCMC is used to assess the inclusion or exclusion of predictors at each spatial location. This section explains the RJMCMC steps mentioned in \ref{algorithm2} and how they're used to find important local variables.\\
\textbf{Initialization:}
\begin{itemize}
    \item Initialize all parameters of the model, i.e., choose initial values for $\gamma_{j}(s)$, $\psi$, $\beta(s)$, $\sigma^2(s)$ and $b$.
    \item Example: Set $\gamma_{j}(s) = 1$ for all $s$ and $j$ (all predictors are included initially).
    \item Set $\psi = 0.5$ assumption that each predictor is equally likely to be included or excluded.
    \item Initialize $\beta(s)$ and $\sigma^2(s)$ based on a simple linear regression with all predictors included.
\item Initialize $b$ based on a suitable initial value (e.g., 10).
\end{itemize}

\textbf{Proposal Step:}
\begin{itemize}
    \item Select a predictor $j$ and a spatial location $s$ randomly.
    \item Propose a change in the model. If $\gamma_{j}(s) = 0$ (predictor $j$ is currently excluded), propose to include predictor $j$ in the model ($\gamma_{j}(s) = 1$). If $\gamma_{j}(s) = 1$ (predictor $j$ is currently included), propose to exclude predictor $j$ from the model ($\gamma_{j}(s) = 0$).
     \item Propose a new value for $b$ by adding a small random change to the current value.
\end{itemize}

\textbf{Calculation of Acceptance Ratio $R$:}
Compute the acceptance ratio as:
\[
R = \frac{p(y|\text{new model}) \times p(\text{new model}) \times q(\text{delete})}{p(y|\text{current model}) \times p(\text{current model}) \times q(\text{add})} \times J
\]

where:

\begin{itemize}
    \item \textbf{Likelihood of Data Given Model $p(y|model)$}: The likelihood function is given by the normal distribution:

    \[
    p(y|\beta(s),x,W(s),\sigma^2(s)) = \pi^{-\frac{n}{2}} \cdot \sigma^{-n}(s) \cdot |W(s)|^{-\frac{1}{2}} \cdot \exp\left(-\frac{1}{2} \sum_{i} w_i^{-1}(s) \cdot (y(s) - x^T(s)\beta(s))^2\right)
    \]

    \item \textbf{Prior Probability of Model $p(model)$}: This term includes the prior distribution of the model parameters $\beta(s)$, $\sigma^2(s)$, and $\psi$.with a Bernoulli prior for $\gamma_{j}(s)$ and a Beta prior for $\psi$,  $\beta(s)$ prior is $N(0, \Sigma_\beta)$ ,$\sigma^2(s)$ prior is $ \text{IGamma}(\alpha_1, \alpha_2)$ and $b$ prior is $\text{uniform}(0,D)$.

    \item \textbf{Proposal Densities $q(add),q(delete)$}: These represent the probability of proposing to add or delete a predictor. In this case, let's assume $q(add)=q(delete)=\frac{1}{p}$, where $p$ is the total number of predictors. This assumes that each predictor is equally likely to be proposed for addition or deletion.

    \item \textbf{Jacobian $J$}: The Jacobian accounts for the change in the dimensionality of the parameter space when adding or deleting predictors. In this case, adding a predictor increases the dimension by one, while deleting reduces it by one. We use the Jacobian to properly adjust our calculations in the RJMCMC algorithm. Specifically, we apply the reverse of the Jacobian that was used when we proposed changing the model. This helps to ensure our probabilities are correct and the Markov chain remains balanced.
\end{itemize}

\textbf{Acceptance/Rejection Step:}
\begin{itemize}
    \item Draw $u \sim \text{Uniform}(0, 1)$. If $u < \min(1, R)$, accept the proposed change in the model; otherwise, keep the current model.
\end{itemize}

\textbf{Repeat:}
\begin{itemize}
 \item Iterate the above steps, allowing the Markov chain to jump between different model spaces, determining the inclusion/exclusion of predictors for each spatial location $i$, and updating the parameters $\beta(s)$, $\sigma^2(s)$, and $b$.
\end{itemize}

\textbf{Post-Processing:}
\begin{itemize}
    \item After running the RJMCMC for a large number of iterations, post-process the Markov chain to obtain posterior distributions of the parameters $\beta(s)$, $\sigma^2(s)$, $\gamma_{j}(s)$, $b$, and $\psi$. These posterior distributions provide estimates of the parameters along with measures of uncertainty.
\end{itemize}
\section{Model Assessment} \label{assessment}
The weighted functions utilised in the model influence the GWR model. Various spatial weighted functions were introduced to be incorporated into the GWR model. To determine the optimal fit for the data, we employed standard evaluation tools, including WAIC \cite{watanabe2010asymptotic}. and  DIC \cite{spiegelhalter2002bayesian}.\\
WAIC is commonly computed from samples of the posterior distribution of interest using a pointwise log predictive density. WAIC is given as:
\begin{align}
\begin{split}\label{eq:10}
    &WAIC=-2 \sum_{i=1}^{n} log \Tilde{p}(y(s)|y)+2V\\   
          & = -2 \sum_{i=1}^{n} log \int p(y(s)|\theta)p(\theta|y) d\theta+2V
\end{split}
\end{align}
where $V=\sum_{i=1}^{i=n} V_i$ is the sum of the posterior variance of the pointwise log-likelihoods.
\begin{equation}
    V_i=Var_{\theta|y} log(p(y(s)|\theta))
\end{equation}
WAIC can be estimated using posterior sample $p(\theta|y)$ to evaluate $\Tilde{p}(y(s)|y)$ and $V$.\\
 To find WAIC for the BGWR model, for each data point $y(s)$: calculate the log-likelihood for each posterior sample $j=1,2,\ldots,M$ using the posterior estimates $\bar{\beta}(s)$, $\bar{W}(i)$, and $\bar{\sigma}^2(s)$:
\begin{equation}
\log f(y(s)|\bar{\beta}(s),X,\bar{W}(i),\bar{\sigma}^2(s)) = \log \left[ \pi^{-n/2} \cdot \bar{\sigma}^{-n} \cdot |\bar{W}|^{-1/2} \cdot \exp\left(-\frac{1}{2} \bar{\sigma}^{-2} (y(s)- X\bar{\beta})^T \cdot \bar{W}^{-1} \cdot (y(s) - X\bar{\beta})\right) \right]
\end{equation}
then calculate the mean log-likelihood over all posterior samples for data point $y(s)$:
\begin{equation}
\overline{\log }f(y(s)) = \frac{1}{M} \sum_{j=1}^{M} \log f(y(s)|\bar{\beta}(s),X,\bar{W}(i),\bar{\sigma}^2(s))
\end{equation}
Take the squared difference between the individual log-likelihood and the mean log-likelihood, and take the average over all the $M$ posterior samples:
\begin{equation}
V_i = \frac{1}{M} \sum_{j=1}^{M} \left(\log f(y(s)|\bar{\beta}(s),X,\bar{W}(i),\bar{\sigma}^2(s)) - \overline{\log }f(y(s)) \right)^2
\end{equation}
Sum up all the individual $V_i$ values to get the total $V$:
\begin{equation}
V = \sum_{i=1}^{n} V_i
\end{equation}
Calculate the log pointwise predictive density for each data point $y(s)$:
\begin{equation}
\log \widetilde{p}(y(s)|y) =  \log f(y(s)|\bar{\beta}(s),X,\bar{W}(i),\bar{\sigma}^2(s))
\end{equation}
finally, calculate the WAIC for the BGWR model using the formula:
\begin{equation}
\text{WAIC} = -2 \sum_{i=1}^{n} \log \widetilde{p}(y(s)|y) + 2V
\end{equation}
where $V$ is the sum of the posterior variances $V_i$ for all data points, and $\log \widetilde{p}(y(s)|y)$ is the log pointwise predictive density for each data point. Smaller WAIC value indicates a better-fitting model.
\\
The DIC is given as :
\begin{equation}
    DIC=Dev(\Bar{\theta}) +2p_D
\end{equation}
In this context, $\theta$ denotes the parameters of interest, and $\bar{\theta}$ represents the posterior mean. The deviance function, denoted as $\text{Dev}(\cdot)$, is utilised, and the effective number of parameters in the model, denoted as $p_D$, is calculated using the equation $p_D = \overline{\text{Dev}}(\theta) - \text{Dev}(\bar{\theta})$. The specific expression for the deviance function is provided in \cite{ma2021bayesian} and is as follows:\\
 \begin{equation}
 \begin{split}
  Dev(\beta(s),W(s),\sigma^2(s))=-2log f(Y|\beta(s),X,W(s),\sigma^2(s))  \\=nlog(2\pi)+log(\sigma^2(s))-log(|W(s)|)+(Y-X\beta(s))^T \sigma^{-2} W(s)(Y-X\beta(s)) 
   \end{split}
 \end{equation}
 The DIC for GWR can be summarised as:
 \begin{equation}
 \begin{split}
     DIC=Dev(\bar{\beta}(s),\bar{W}(i),\bar{\sigma}^2(s))+2p_D\\
     = 2 \overline{Dev}(\beta(s),W(s),\sigma^2(s))-Dev(\bar{\beta}(s),\bar{W}(i),\bar{\sigma}^2(s))
      \end{split}
 \end{equation}
The quantities $\bar{\beta}(s)$, $\bar{W}(i)$, and $\bar{\sigma}^2(s)$ correspond to the posterior estimates derived from the MCMC results similar to before. A smaller DIC value is a more favourable model \cite{ma2021geographically}.

\section{DPMM Cluster Configuration Process} \label{DPMMapp}
In this section, we provided a further explanation for the two-stage cluster configuration that we developed to find the final cluster labelling for the DPMM in the case of the mode method.

\textbf{ Stage 1: Within-Sample Cluster Configuration}

Given:
\begin{itemize}
    \item \(N\): total number of regions
    \item \(M\): number of MCMC samples obtained from BGWR (here, \(M = 500\))
      \item \(Q\): number of MCMC iterations for DPMM for each sample $i$ (here, \(Q = 2000\))
    \item \(C_i\): cluster configuration for the \({ith}\) sample
\end{itemize}

For each sample \(i\) from 1 to \(M\):
\begin{enumerate}
    \item Perform MCMC analysis on the DPMM and determine the cluster assignment for each region.
    \item Based on the given regions, we can find the cluster configuration, \(C_i\), using the mode method.
    \item Arrange the cluster configuration derived from each sample$i$ across $Q$ iteration and put $NA$ if the value does not exist. Each sample $i$ may give a different number of clusters.
\end{enumerate}

Mathematically, for the mode method, the cluster assignment for a region \(j\) during the \({ith}\) iteration, \(c_{i, j}\), is:
\[ c_{i,j} = \text{mode} \left( p(x_j | C_i) \right) \]
where \( p(x_j | C_i) \) represents the probability of region \(N_i\) belonging to each cluster in configuration \(C_i\).

\textbf{Stage 2: Across-Sample Cluster Configuration}

Given the set of cluster configurations from the first stage:
\[ C = \begin{bmatrix}
c_{1,1} & c_{1,2} & \dots & c_{1,j} \\
c_{2,1} & c_{2,2} & \dots & c_{2,j} \\
\vdots & \vdots & \ddots & \vdots \\
c_{M,1} & c_{M,2} & \dots & c_{M,j} \\
\end{bmatrix} \]

Where:
\begin{itemize}
    \item \(c_{i,j}\) is the cluster assignment for region \(j\) for sample \(i\).
    \item If the number of clusters during sample \(i\) is less than \(M_{\text{max}}\), some \(c_{i,j}\) values in that row will be \texttt{NA}.
\end{itemize}
\begin{enumerate}
    \item For each region \(j\), consider its column in matrix \(C\).
    \item Calculate the mode across all entries in that column, ignoring the \texttt{NA} values.
\end{enumerate}

The final cluster assignment for each region \(j\), \(c_{\text{final},j}\), is:
\[ c_{\text{final},j} = \text{mode}\left( \{c_{1,j}, c_{2,j}, \dots, c_{i,j}\} \right) \]

Where:
\begin{itemize}
    \item \(c_{\text{final},j}\) is the most frequently assigned cluster for region \(j\) across all MCMC 500 samples.
    \item The mode function will exclude \texttt{NA} values.
\end{itemize}

Using the matrix \(C\), we construct the final cluster assignment vector \(C_{\text{final}}\) as:
\[ C_{\text{final}} = \begin{bmatrix}
c_{\text{final},1} \\
c_{\text{final},2} \\
\vdots \\
c_{\text{final},j} \\
\end{bmatrix} \]

Where each entry \(c_{\text{final},j}\) corresponds to the final cluster assignment for each region \(j\). To handle variations in cluster numbers across iterations, the use of \texttt{NA} simplifies the mode calculation by indicating a lack of assignment.
\section{Simulated Data Further Analysis}   \label{sim_further_test}
In this section, we present additional results from the analysis of simulated data for the three settings across 100 replicates and report the average performance within these replicates. Our main focus is on the spatial distributions of the clusters obtained using our proposed algorithm. The final cluster configuration is determined using both Dahl's method and the mode method as the assignment criteria.

\begin{figure}[ht!]
    \centering
    \includegraphics[scale=0.4]{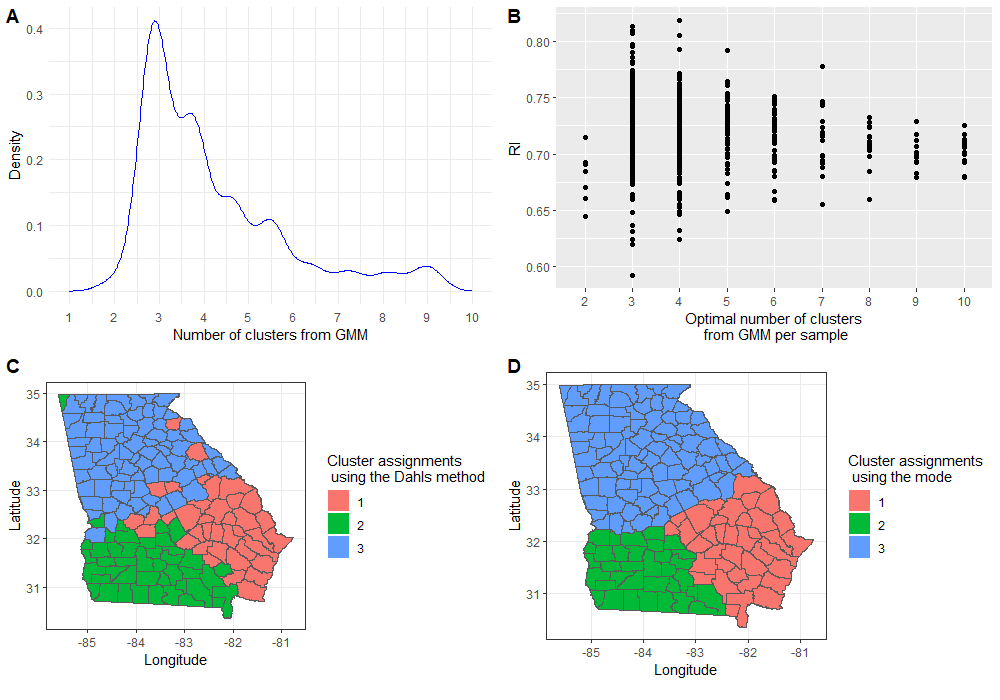}
    \caption{The optimal number of clusters obtained from the GMM using BIC from 500 samples from the posterior, for Setting 2, along with the cluster configuration on the spatial map using the proposed algorithm }
    \label{fig:enter-label3}
\end{figure}

\begin{figure}[ht!]
    \centering
    \includegraphics[scale=0.4]{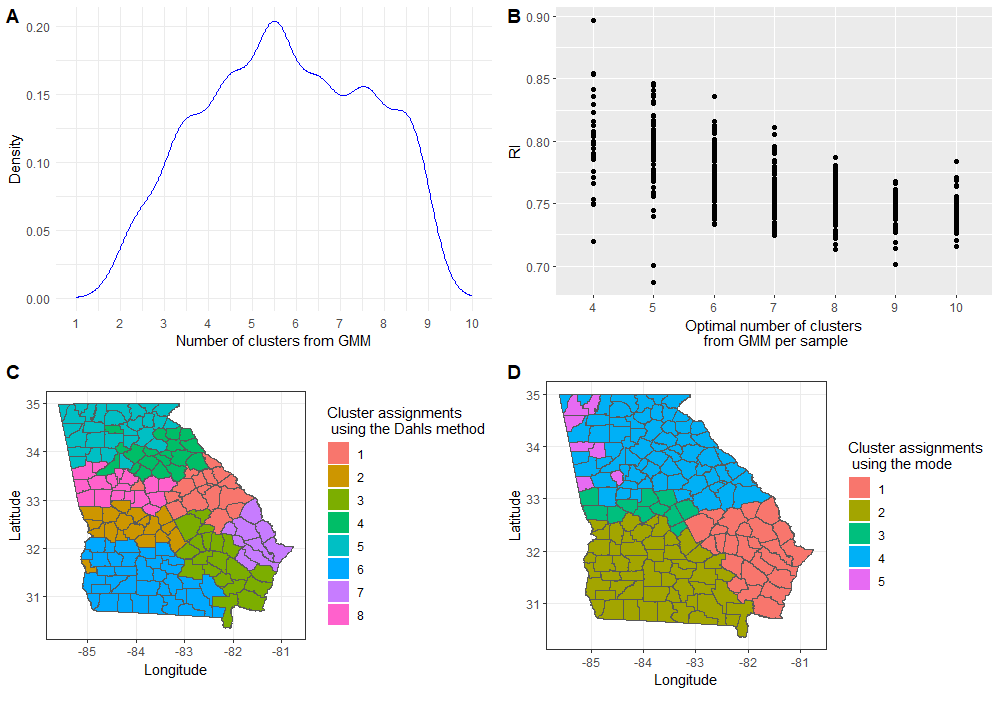}
    \caption{The optimal number of clusters obtained from the GMM using BIC from 500 samples from the posterior, for Setting 3, along with the cluster configuration on the spatial map using the proposed algorithm. }
    \label{fig:enter-label4}
     \end{figure}

Additionally, in this section, we present an empirical density plot showcasing the optimal number of clusters obtained from each sample of the BGWR coefficients posteriors. To ensure accuracy, we considered approximately 500 samples from the posterior of the beta parameters obtained from BGWR without replacement during the evaluation of our algorithm. For the simulated data, we ran our algorithm for 5000 iterations, with 2000 burn-in iterations.
\begin{figure}[ht!]
    \centering
    \includegraphics[scale=0.4]{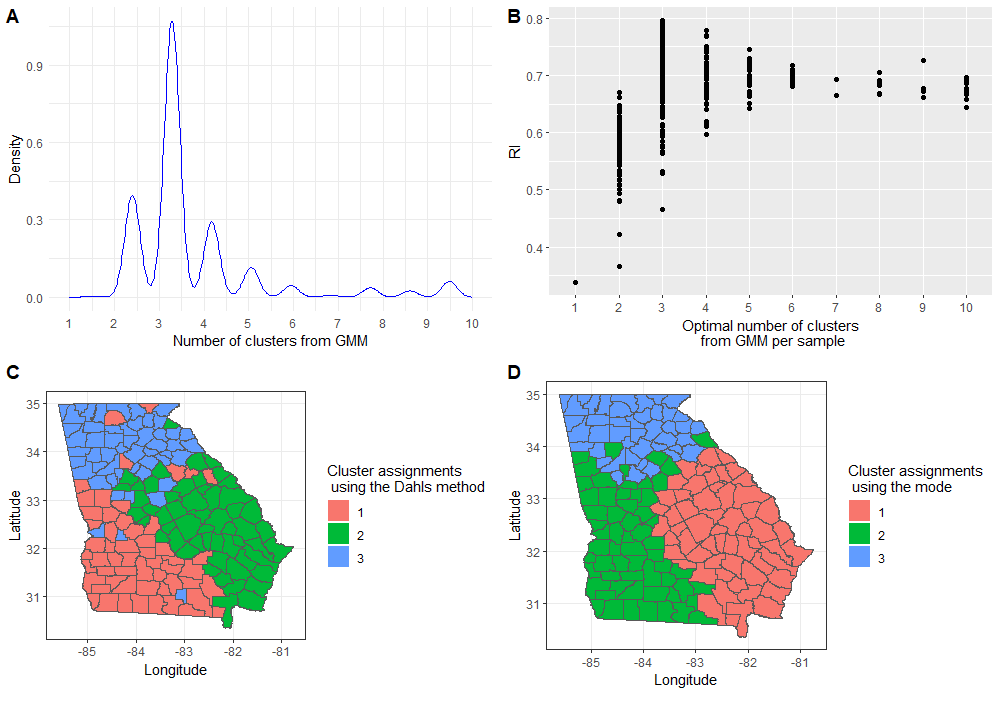}
    \caption{The optimal number of clusters obtained from the GMM using BIC from 500 samples from the posterior, for Setting 1, along with the cluster configuration on the spatial map using the proposed algorithm }
    \label{fig:enter-label2}
    \end{figure}
Furthermore, Figure \ref{fig:enter-label3} depicts the spatial distribution obtained from the proposed algorithm using the GMM cluster method. Notably, our algorithm consistently identifies the optimal number of clusters as 3 across the mode method and Dahl's method. Noticeably, in this setting, the range of the Rand index (RI) is higher compared to setting 1.
Compared to Figure \ref{fig:enter-label2} and Figure \ref{fig:enter-label3}, Figure \ref{fig:enter-label4} demonstrates higher accuracy (RI) with the true labels. The Mode method and Dahls' method generate more clusters. As anticipated, increasing the signal strength leads to improved clustering accuracy. However, the proposed approach exhibits a tendency for over-clustering. Nevertheless, this tendency diminishes as the signal strength increases.

We compared two models for grouping data, DPMM and GMM, in three different situations. Notably, the DPMM outperformed the GMM by creating more clusters, especially smaller ones. This pattern persisted consistently across all three scenarios, demonstrating the DPMM's remarkable ability to uncover structures in the data. The DPMM's adaptive nature allowed it to identify complex patterns, making it a powerful tool for revealing hidden insights in complex datasets.
    \begin{figure}[ht!]
        \centering
        \includegraphics[scale=0.35]{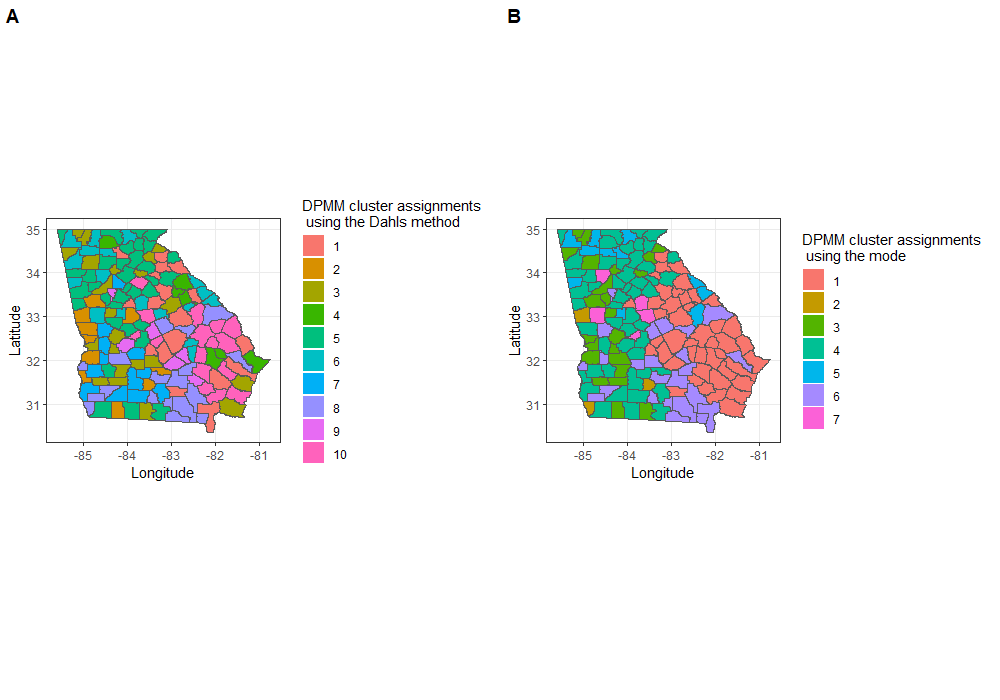}
        \caption{The spatial distribution of the clusters obtained from setting 1 using DPMM.}
        \centering
        \includegraphics[scale=0.35]{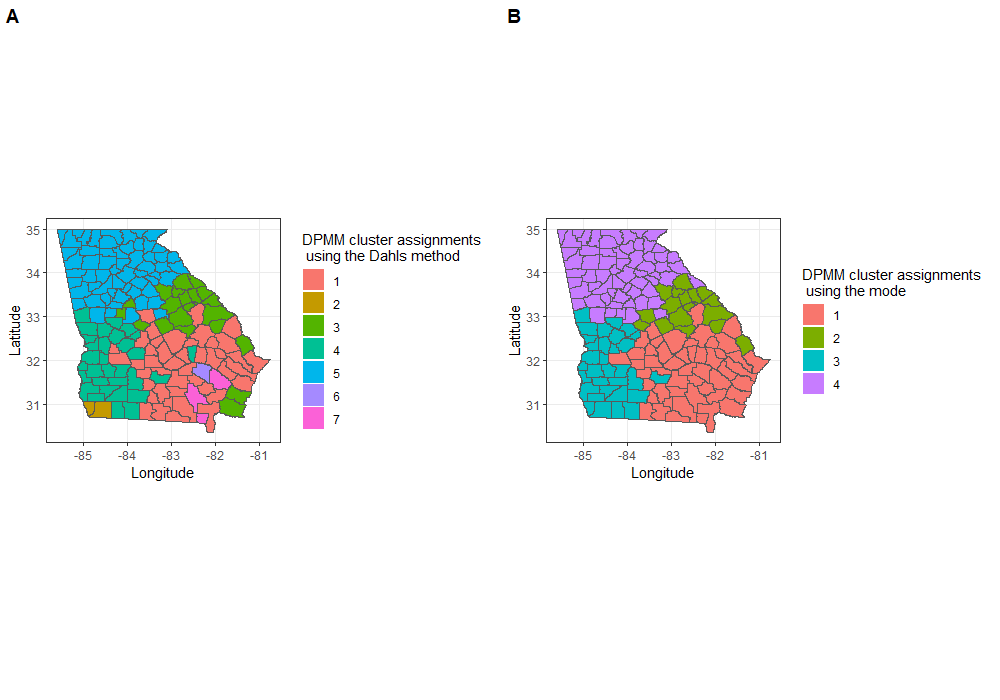}
        \caption{The spatial distribution of the clusters obtained from setting 2 using DPMM.}
        \centering
        \includegraphics[scale=0.35]{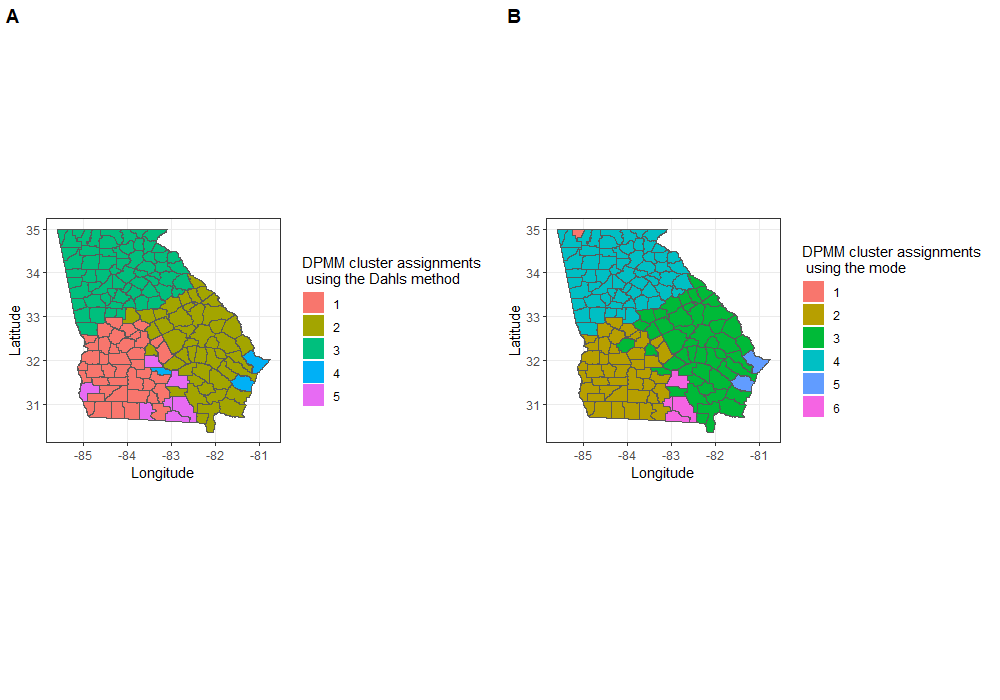}
        \caption{The spatial distribution of the clusters obtained from setting 3 using DPMM.}
        \label{fig:enter-label7}   
\end{figure}
\newpage
\section{Case Study Further Analysis} \label{real_further_test}
\begin{figure}[ht!]
    \centering
    \includegraphics[scale=0.4]{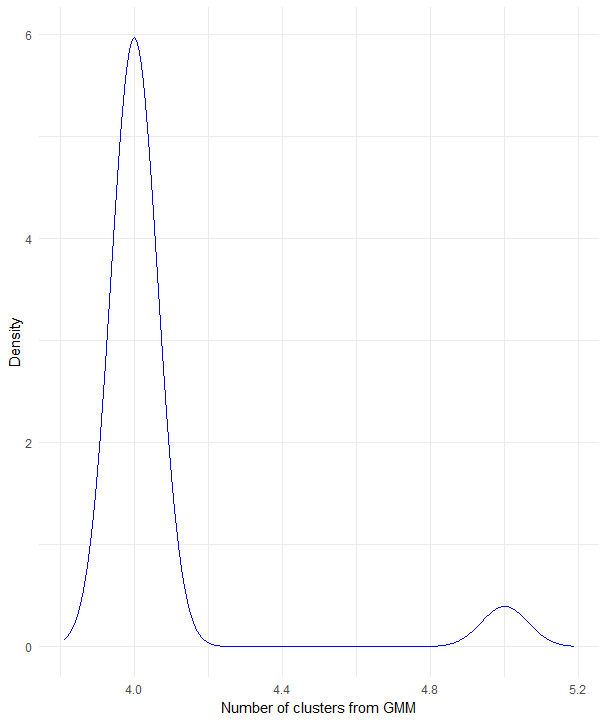}
    \includegraphics[scale=0.4]{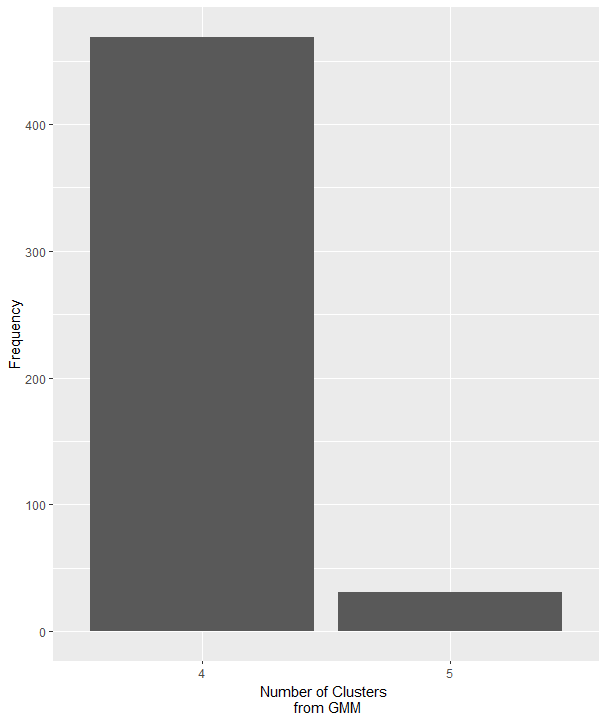}
    \caption{Optimal cluster distribution per sample. }
    \label{fig:enter-label5}
\end{figure}

The optimal number of cluster derived from 500 samples with the GMM method and assessed using BIC, is depicted in Figure \ref{fig:enter-label5}.
The provided plots illustrate the probability of each region belonging to various clusters. Since we have 526 regions to visualize on the heatmap, we have presented this information across six figures, with the first five  figures displaying the probabilities for 100 regions. Similarly, we displayed the probability of each region being assigned to one of the 12 clusters obtained from the DPMM in a heatmap.
\begin{figure}[ht!]
    \centering
    \includegraphics[scale=0.35]{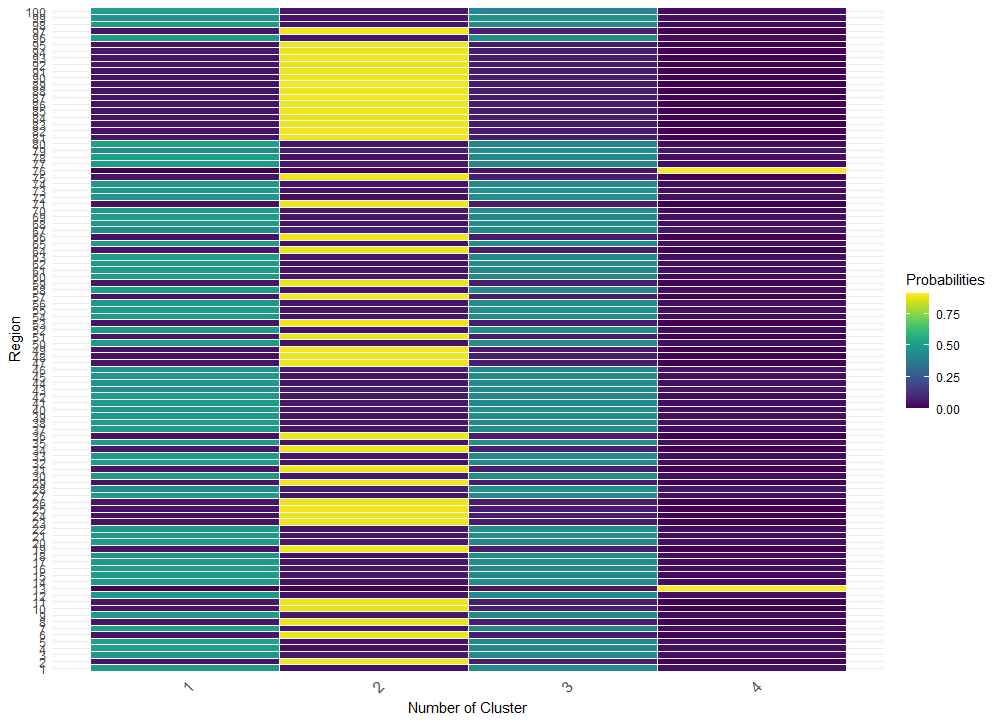}
    \caption{The probabilities of the first 100 regions to belong to each cluster from 1 to 4 for GMM.}
\end{figure}
\begin{figure}[ht!]
    \centering
    \includegraphics[scale=0.35]{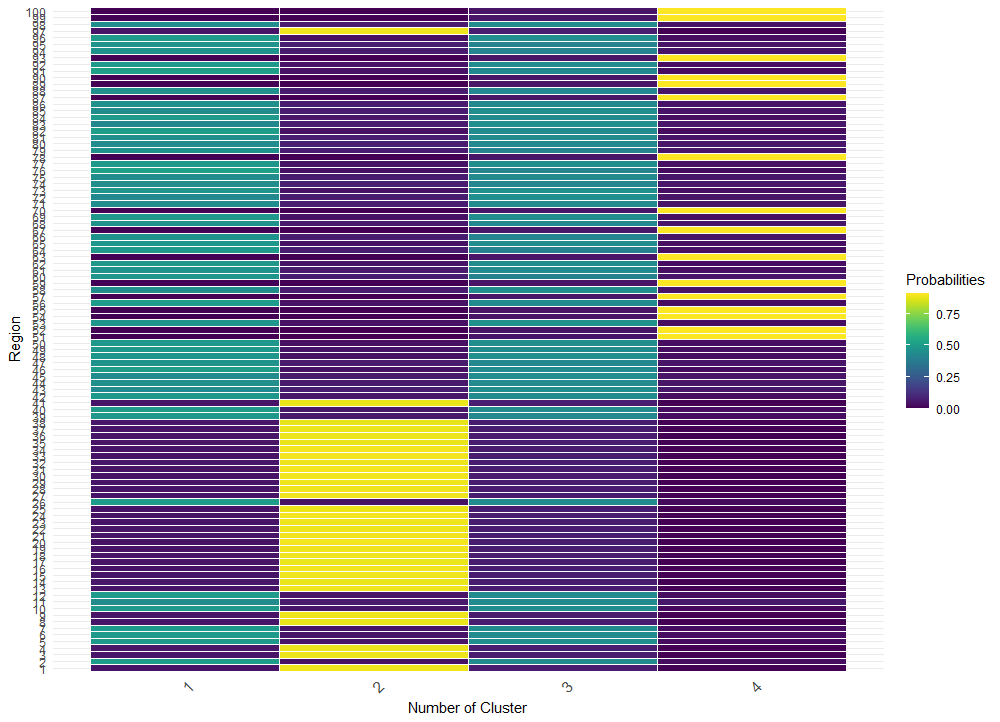}
    \caption{The probabilities of the second 100 regions to belong to each cluster from 1 to 4 for GMM.}
\end{figure}
\begin{figure}[ht!]
    \centering
    \includegraphics[scale=0.35]{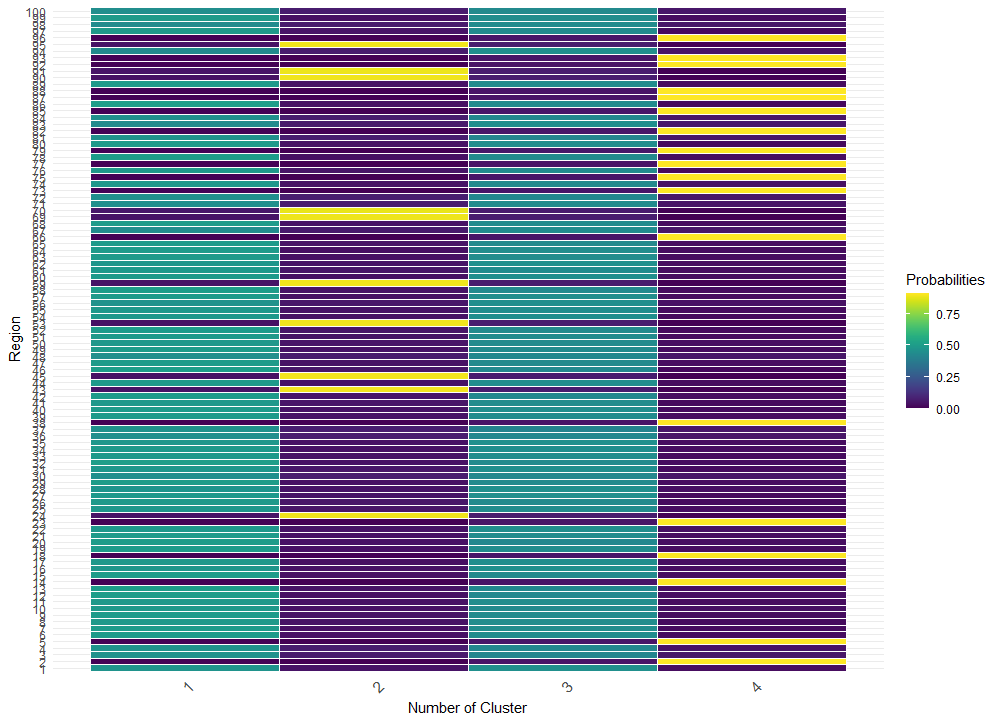}
    \caption{The probabilities of the third 100 regions to belong to each cluster from 1 to 4 for GMM.}
\end{figure}
\begin{figure}[ht!]
    \centering
    \includegraphics[scale=0.4]{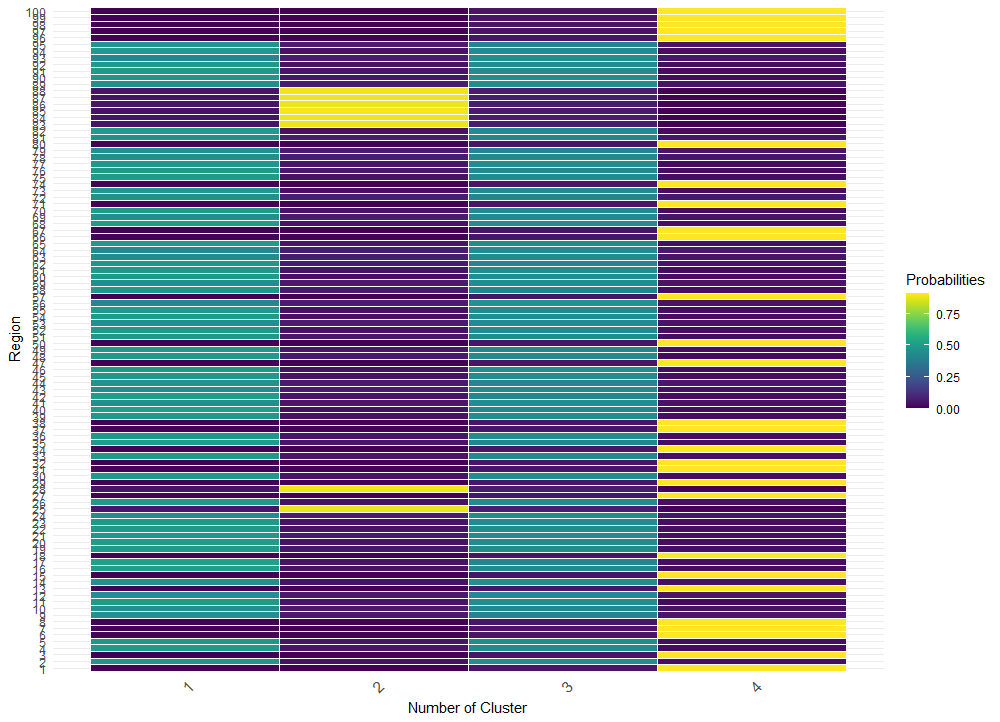}
    \caption{The probabilities of the fourth 100 regions to belong to each cluster from 1 to 4 for GMM.}
\end{figure}
\begin{figure}[ht!]
    \centering
    \includegraphics[scale=0.4]{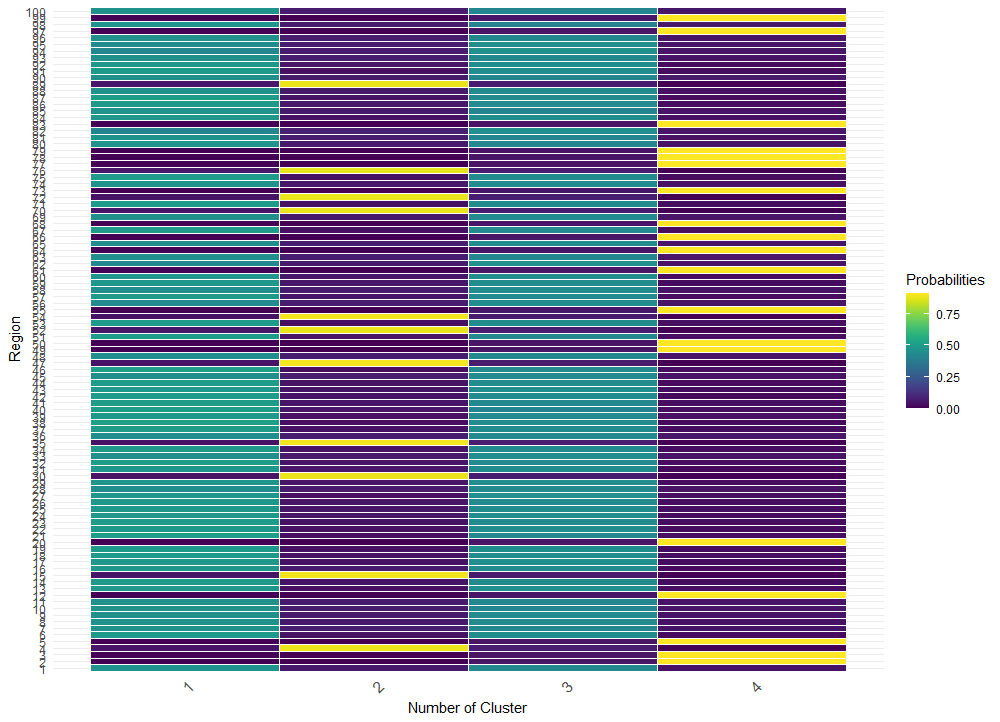}
    \caption{The probabilities of the fifth 100 regions to belong to each cluster from 1 to 4 for GMM.}
\end{figure}
\begin{figure}[ht!]
    \centering
    \includegraphics[scale=0.4]{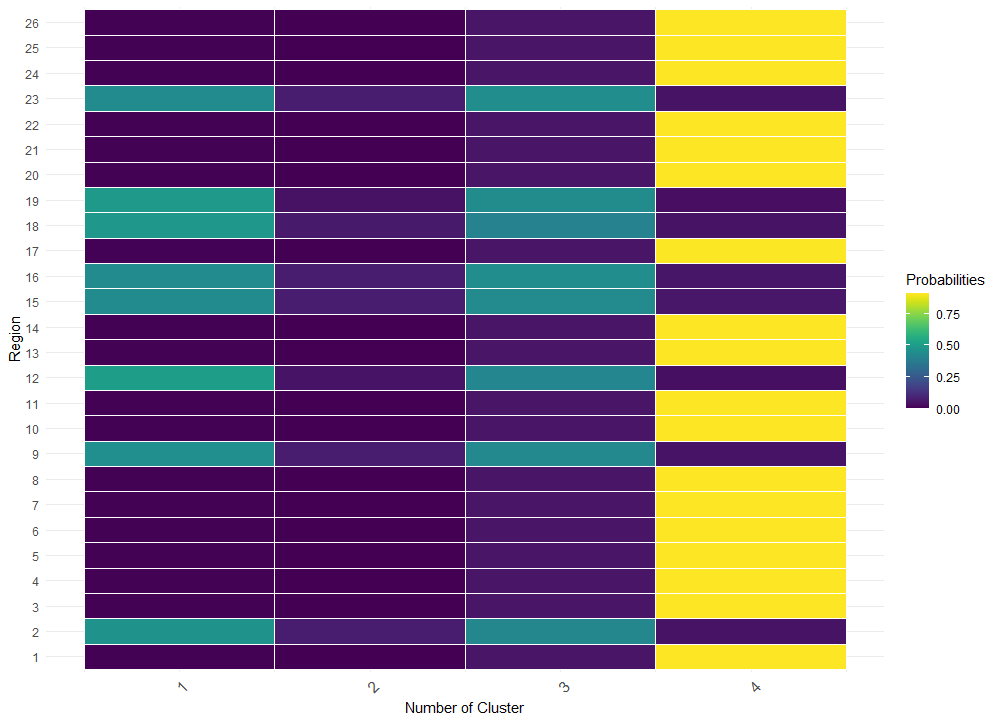}
    \caption{The probabilities of the last 26 regions to belong to each cluster from 1 to 4 for GMM.}
\end{figure}

\begin{figure}[ht!]
    \centering
    \includegraphics[scale=0.5]{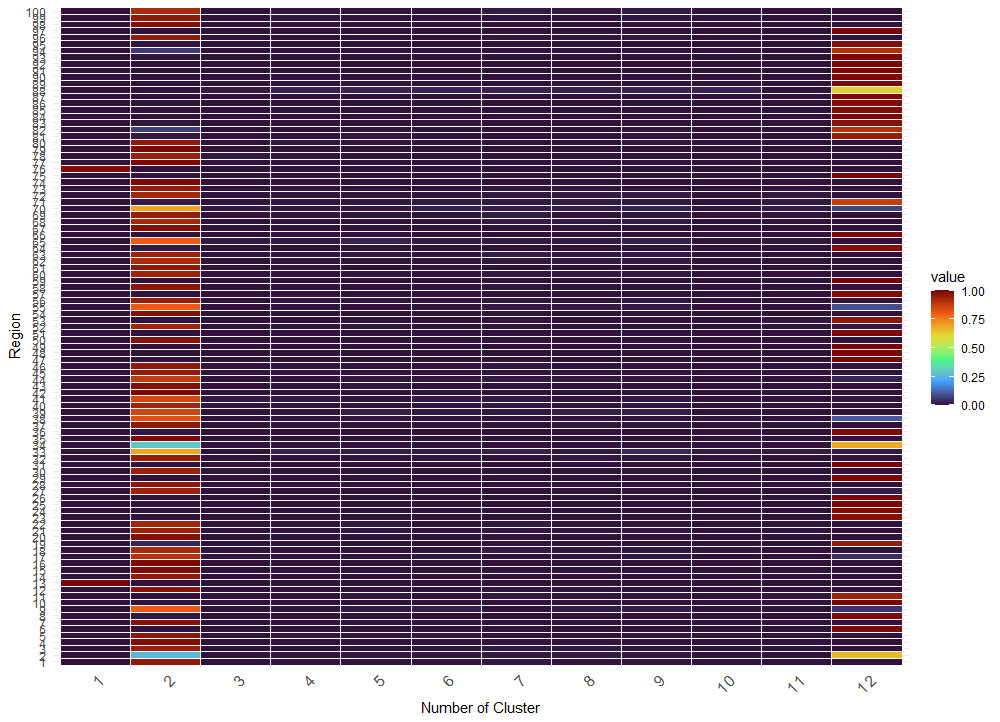}
    \caption{The probabilities of the first 100 regions to belong to each cluster from 1 to 12 for DPMM.}
    \centering
    \includegraphics[scale=0.5]{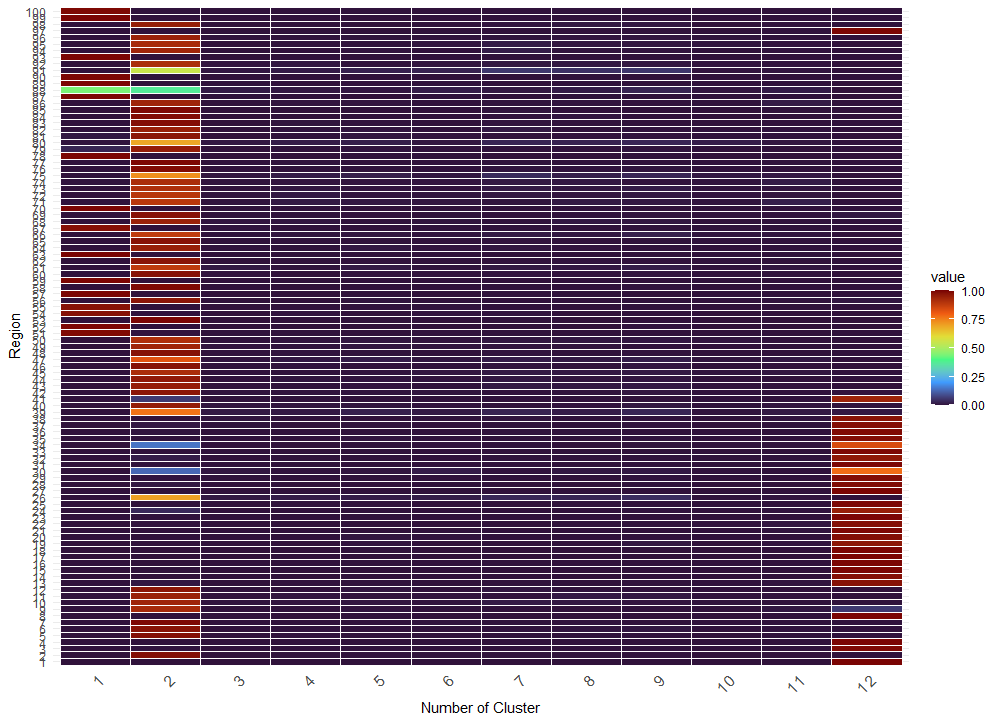}
    \caption{The probabilities of the second 100 regions to belong to each cluster from 1 to 12 for DPMM.}
\end{figure}

\begin{figure}[ht!]
    \centering
    \includegraphics[scale=0.5]{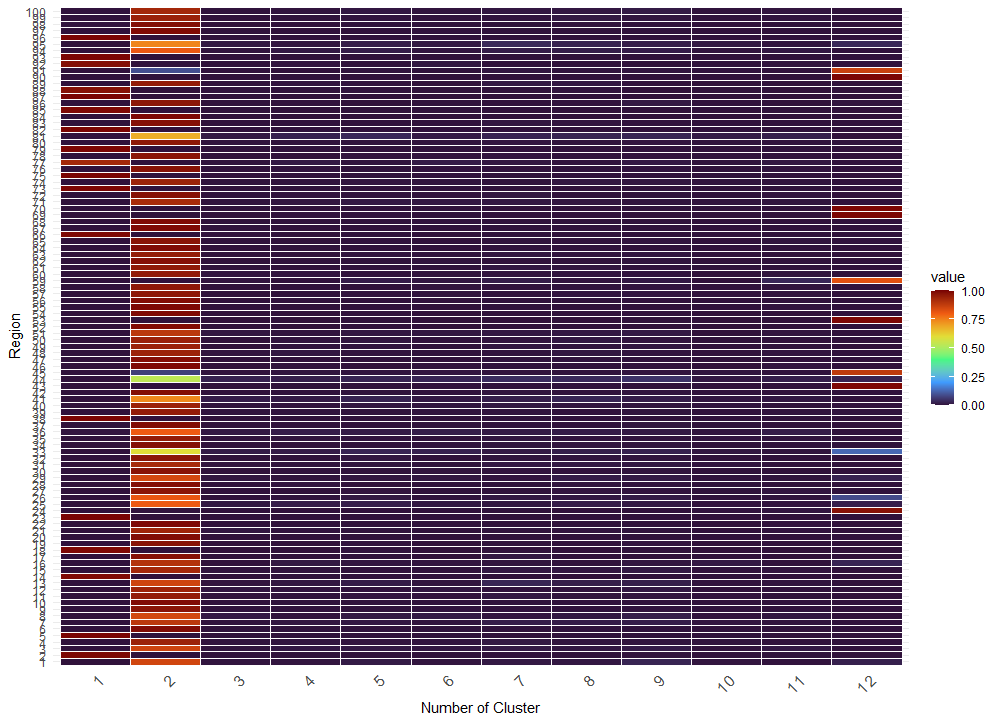}
    \caption{The probabilities of the third 100 regions to belong to each cluster from 1 to 12 for DPMM.}
    \centering
    \includegraphics[scale=0.5]{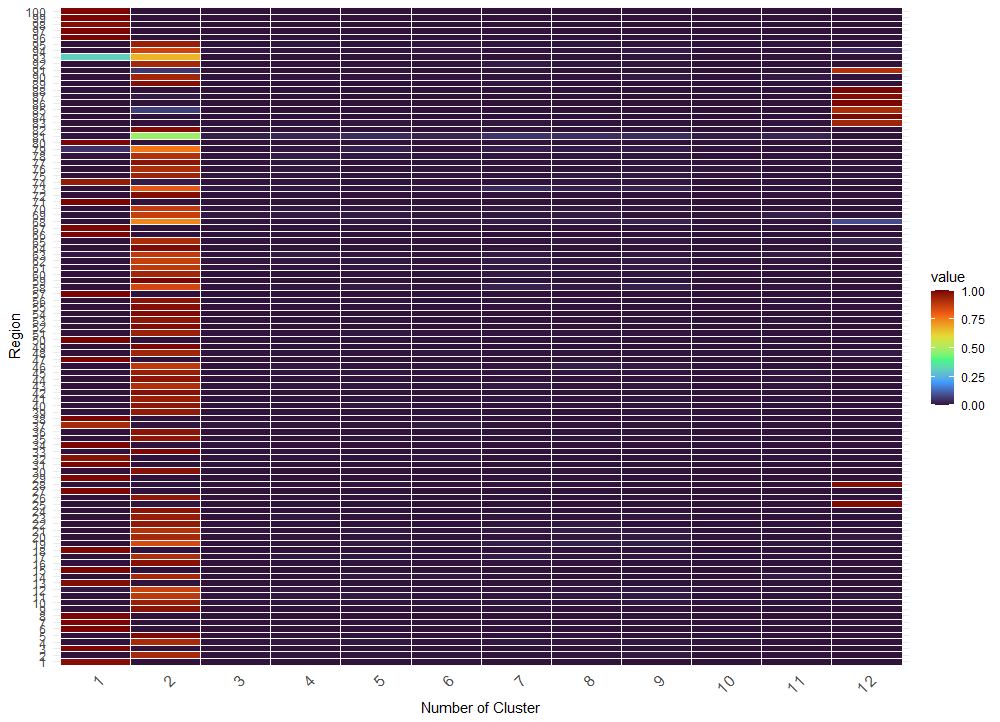}
    \caption{The probabilities of the fourth 100 regions to belong to each cluster from 1 to 12 for DPMM.}
\end{figure}
\begin{figure}[ht!]
    \centering
    \includegraphics[scale=0.5]{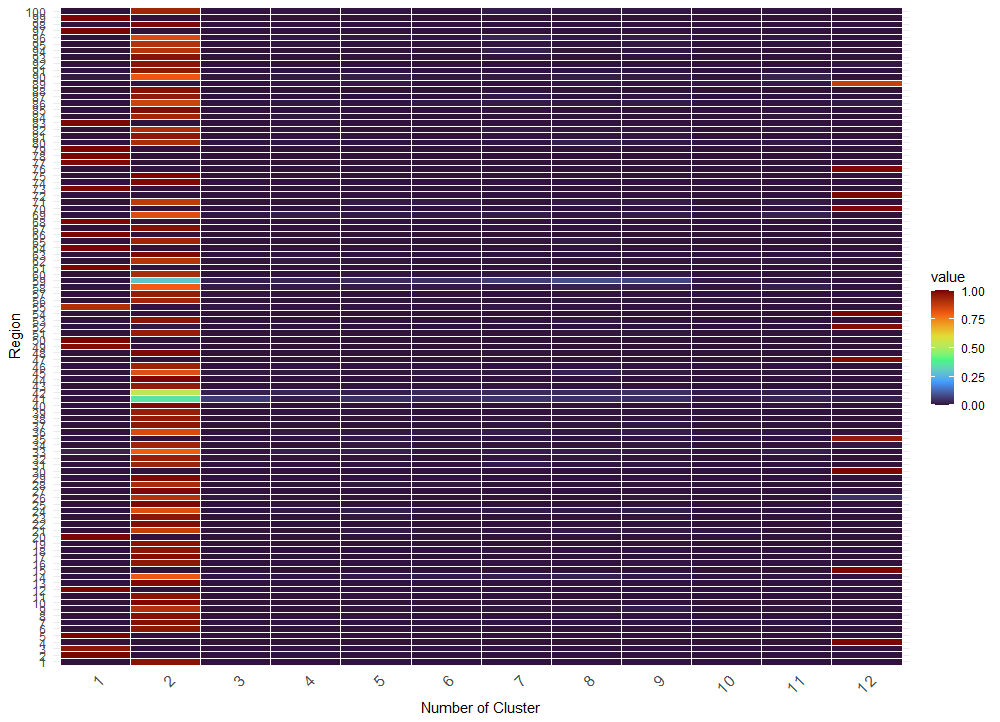}
    \caption{The probabilities of the fifth 100 regions to belong to each cluster from 1 to 12 for DPMM.}

    \centering
    \includegraphics[scale=0.5]{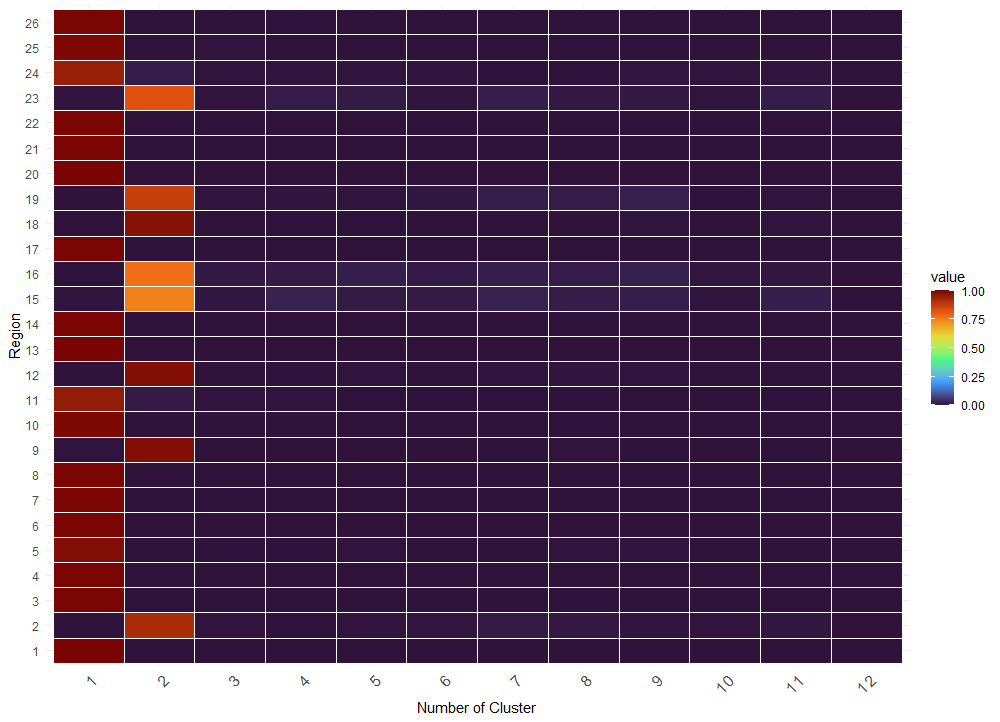}
    \caption{The probabilities of the last 26 regions to belong to each cluster from 1 to 12 for DPMM.}
\end{figure}

\end{appendices}


\end{document}